\documentclass[11pt,a4paper]{article}
\usepackage[margin=2.6cm]{geometry}
\usepackage{amsmath,amssymb,amsthm,mathtools}
\usepackage{booktabs}
\usepackage{enumitem}
\usepackage{tikz-cd}
\usepackage[colorlinks=true,linkcolor=blue!50!black,citecolor=green!40!black,urlcolor=blue!50!black]{hyperref}
\usepackage{cleveref}
\usepackage{xcolor}

\numberwithin{equation}{section}
\newtheorem{theorem}{Theorem}[section]
\newtheorem{proposition}[theorem]{Proposition}

\theoremstyle{definition}
\newtheorem{definition}[theorem]{Definition}
\newtheorem{example}[theorem]{Example}
\theoremstyle{remark}
\newtheorem{remark}[theorem]{Remark}

\newtheorem{uefn}[]{Principle} 
\newenvironment{principle}{\begin{uefn}}{\end{uefn}}  

\newcommand{\R}{\mathbb{R}}
\newcommand{\C}{\mathbb{C}}
\newcommand{\Mot}{\mathcal{M}}
\newcommand{\Dbl}{\mathbb{D}_J}
\newcommand{\Pol}{\mathcal{P}_J}
\newcommand{\Sp}{\mathrm{Sp}}
\renewcommand{\sp}{\mathfrak{sp}}
\newcommand{\Jp}{\mathcal{J}_+(V,\omega)}
\newcommand{\st}{\star_{J,\lambda}}
\newcommand{\stM}{\star_{\mathrm M,\lambda}}
\newcommand{\vol}{\mu_\omega}
\newcommand{\Koop}{\mathfrak{L}}
\newcommand{\wb}{\bar w}
\newcommand{\T}{\mathsf{T}}
\newcommand{\Fock}{\mathcal{F}_{J,\lambda}}

\title{Covariance Cells:\\[2pt]
\large Classical Probability Amplitudes, Positive Polarizations,\\ and Star Products of Gaussian Resolution}
\author{Carlos Alberto Alcalde\\[2pt]
\small NXP Semiconductors, Grenoble, France \quad \texttt{alcaldecarlos09@gmail.com}\\
\small ORCID 0000-0002-4667-8154}
\date{}

\begin{document}
\maketitle

\begin{abstract} In classical statistics, a Gaussian preparation is not a point of phase space but a \emph{covariance cell}: an ellipsoid whose size, shape, and orientation are determined by a positive covariance tensor. We show that, on a linear symplectic space \((V,\omega)\), balanced covariance cells are naturally identified with positive polarizations. Here \emph{balanced} means that the preparation has a single common symplectic resolution scale. After separating this scale, compatible complex structures \(J\), positive Lagrangian subspaces, Kähler metrics \(g_J\), and normalized covariance tensors \(\Sigma_J=g_J^{-1}\) are four descriptions of the same object, and their parameter space is the Siegel space \(\Sp(V,\omega)/\mathrm U(V,\omega,J)\). Working in the Koopman--von~Neumann representation, where probability measures are represented by square-root amplitudes and physical observables act as commuting multiplication operators, we establish the following results. First, along a reference motion, a symplectic connection expresses the linearized dynamics as a time-dependent quadratic Koopman evolution on one fixed tangent fibre; Gaussian covariances are transported by \[ \Sigma_t=S_t\Sigma_0S_t^{\T}. \] Second, every balanced cell determines a separated star product \(\star_{J,\lambda}\) satisfying \[ [\wb^{\bar a},z^b]_\star = \lambda\Sigma_J^{\bar ab}. \] This product is the Moyal product transformed by Gaussian smoothing at the resolution of the cell. Third, a quadratic Hamiltonian acts through its classical Poisson derivation inside a fixed cell algebra exactly when its flow preserves the cell. Cell-changing squeezes are instead implemented by isomorphisms between the algebras of the initial and transported cells. Fourth, Fronsdal's right star-polarization equations select, by associativity, a left module whose cyclic element is the Gaussian density of the cell. Its completion is the corresponding Fock space, and Koopman transport maps the Fock space of one cell unitarily onto that of the transported cell. No Planck constant is introduced: \(\lambda\) is a classical resolution scale, and measured observables remain commutative. The construction identifies the covariance cell, positive polarization, Gaussian smoothing, cell star algebra, and star-polarized Fock module as different realizations of one classical statistical geometry. \end{abstract}



\vspace{4pt}
\begin{flushright}\small\itshape To the memory of Chris and Moshe\end{flushright}

\tableofcontents

\section{Introduction}\label{sec:intro} \subsection{The question} Ask an experimentalist where a particle is, and the complete statistical answer is not a point. It is a region, with a given size and orientation, in which the particle is found with a specified confidence. In phase space, the simplest such region is a Gaussian \emph{covariance cell}: an ellipsoid defined by a level set of \[ x\longmapsto x^{\T}\Sigma^{-1}x, \] where \(\Sigma\) is a positive covariance tensor. Hamiltonian mechanics, by contrast, is formulated in terms of points and their trajectories. This paper asks what geometry, what algebra of observables, and what notion of state emerge when the covariance cell, rather than its centre alone, is taken as the basic object of classical statistical mechanics. The question brings together several established viewpoints. Operational approaches take as primitive the statistical pairing between preparations and observations \cite{Holevo2011,BuschGrabowskiLahti1997}. Souriau's \emph{espace des mouvements} replaces the instantaneous point by the complete motion~\cite{Souriau1970}. Symplectic transport may be represented by a Lagrangian relation in a doubled space \cite{GuilleminSternberg1984}, while deformation theory separates equations of motion from initial conditions and associates them, respectively, with observables and states~\cite{BFFLS1978}. The purpose of this paper is to assemble these ideas into a framework in which the width and orientation of a preparation are transported with the same geometric care as its centre. \subsection{Preparations, observations, and transport} The empirical content of a physical theory lies in the statistical relation \begin{equation}\label{eq:operational-pairing} (s,A)\longmapsto\sigma(A\mid s) \end{equation} between a preparation \(s\) and an observation \(A\). States and observables are operationally dual: a state specifies how statistics vary with the observation, while an observable specifies how they vary with the preparation. \begin{principle}[Operational pairing principle] \label{pr:operational-pairing} The primitive empirical object is the statistical pairing between preparations and observations. States and observables are defined through this pairing, and physical predictions are obtained from the corresponding statistical assignment. \end{principle} In ordinary probability theory, a preparation is represented by a probability measure \(\mu\) on a measurable space \((\Omega,\mathcal F)\), an observable by a measurable function \(f\), and their pairing by \[ \omega_\mu(f)=\int_\Omega f\,d\mu. \] The same state admits a Hilbert-space realization by a square-root amplitude~\cite{Accardi1976}: \[ \Psi_\mu=\sqrt{\mu}, \qquad \omega_\mu(f) = \langle\Psi_\mu,M_f\Psi_\mu\rangle, \qquad M_f\Psi=f\Psi. \] This rewriting does not introduce a new physical state. It represents the same probability measure in a linear space. In particular, the physical observables in the classical Koopman realization remain commuting multiplication operators. Dynamics adds a transport law to the operational pairing. Following Souriau, we regard a complete motion as the underlying dynamical object, while its value at a given time is one instantaneous realization~\cite{Souriau1970}. A Hamiltonian flow \(\Phi_t:P\to P\) transports observables and amplitudes by Koopman pullback: \[ U_tf=f\circ\Phi_t^{-1}, \qquad U_t\psi=\psi\circ\Phi_t^{-1}. \] When \(\Phi_t\) preserves Liouville volume, \(U_t\) is unitary. This is the classical Hilbert-space formulation introduced by Koopman and von Neumann~\cite{Koopman1931,vonNeumann1932}. Hamiltonian transport also has a relational geometry. Its graph \[ \Gamma_t = \{(x,\Phi_t(x)):x\in P\} \subset P^-\times P \] is Lagrangian for the doubled symplectic form \[ \Omega_{\mathbb D} = -\pi_1^*\omega+\pi_2^*\omega \] \cite{GuilleminSternberg1984}. The doubled space therefore represents transport between initial and final realizations, rather than merely an enlargement of phase space. \subsection{Covariance cells and positive polarizations} We study preparations localized near a reference motion. A symplectic connection identifies its time-dependent tangent spaces with one fixed symplectic vector space \((V,\omega)\), following the linearization framework of Marsden, Ratiu, and Raugel \cite{MarsdenRatiuRaugel1991}. In this fixed fibre, a localized preparation is described by a Gaussian covariance tensor. Every compatible complex structure \(J\) on \(V\) determines a positive metric and an inverse covariance tensor, \[ g_J(u,v)=\omega(u,Jv), \qquad \Sigma_J=g_J^{-1}. \] Conversely, a positive covariance \(\Sigma\) has the form \[ \Sigma=\lambda\Sigma_J \] for a unique compatible \(J\) precisely when all its Williamson symplectic eigenvalues are equal to \(\lambda\) (\cref{prop:balanced}). We call these covariances \emph{balanced cells}. Balanced covariance cells are therefore equivalent to positive polarizations. At fixed scale, their parameter space is the Siegel symmetric space \[ \Jp \simeq \Sp(V,\omega)/\mathrm U(V,\omega,J_0), \] whose underlying normal-form and symplectic geometry originate in \cite{Williamson1936,Siegel1943}. The Cartan splitting \[ \sp(V,\omega) = \mathfrak k_J\oplus\mathfrak p_J \] then separates transformations that preserve the cell from those that squeeze it. \subsection{Main results} The paper develops the consequences of identifying covariance cells with positive polarizations. \begin{enumerate}[label=\textbf{C\arabic*.}, leftmargin=2.2em] \item \textbf{Classical amplitudes and fixed-fibre linearization.} Probability measures are represented by square-root amplitudes, and their Hamiltonian transport by the Koopman operator. A symplectic connection transfers the first variation along one motion to a fixed fibre, where it is generated by a time-dependent quadratic Hamiltonian (\cref{prop:linearKoopman}). Covariance is pushed forward, whereas precision is pulled back. \item \textbf{Cells as positive polarizations.} Balanced covariances, compatible complex structures, and positive Lagrangian polarizations are shown to be equivalent (\cref{prop:balanced}). Their common parameter space is \(\Jp\), with \(\mathfrak k_J\) describing cell-preserving rotations and \(\mathfrak p_J\) describing cell-changing squeezes. \item \textbf{The star product of a cell.} Each balanced cell determines a separated star product \(\star_{J,\lambda}\). Its antisymmetric contraction is fixed by the symplectic form, while its symmetric contraction is the covariance tensor. The product is related to the Moyal product by a Gaussian heat operator whose square is convolution with the cell density (\cref{thm:smoothing}). Thus the symmetric ordering parameter is identified with the covariance of a classical preparation. \item \textbf{Rotations and squeezes.} For a quadratic Hamiltonian \(H_A\), \[ \frac{i}{\lambda}\operatorname{ad}_{\star}(H_A) = \{H_A,\cdot\} \] exactly if and only if \(A\) preserves the cell (\cref{thm:inner}). Rotations therefore act classically within a fixed cell algebra, while squeezes transport between the algebras associated with different cells. \item \textbf{The Fock module of a cell.} Following Fronsdal's star-polarization construction \cite{Fronsdal1979}, one-sided star equations select a left module whose cyclic element is the Gaussian density of the cell. Its completion is the Bargmann--Fock space~\cite{Bargmann1961}, and it embeds isometrically into the classical Koopman amplitude space (\cref{thm:fock}). Koopman transport carries the Fock space of one cell onto that of the transported cell (\cref{thm:koopmanfock}). \end{enumerate} The resulting correspondence is \[ \begin{aligned} \text{statistical preparation} &\longleftrightarrow \text{covariance cell} \longleftrightarrow \text{positive polarization} \\ &\longleftrightarrow \text{cell star algebra} \longleftrightarrow \text{star-polarized Fock module}. \end{aligned} \] These are different realizations of one statistical and symplectic geometry, transported together by the symplectic group. \subsection{Scope and interpretation} No Planck constant is introduced. The parameter \(\lambda\) is the common Williamson scale of a balanced covariance cell. Relative to the chosen symplectic normalization, it has the dimensions of an action, or phase-space area. It is a classical resolution parameter, and \[ \lambda\longrightarrow0 \] is the singular limit of infinitely sharp preparations. In the Koopman realization, physical observables remain commuting multiplication operators. The noncommutative star product organizes the composition of their finite-resolution representatives; it does not turn them into noncommuting physical measurements. Every Gaussian used below is a classical confidence density, and every Hilbert-space formula is a square-root realization of an ordinary probability integral. The star algebra is autonomous: its product, dynamics, and algebraic spectral theory do not require the prior introduction of a Hilbert space. Fronsdal's star polarization is used subsequently to extract a linear module useful for applications. \subsection{Relation to previous work} The paper continues \cite{AlcaldeTyagi2023a,AlcaldeTyagi2023b,Alcalde2025}. In~\cite{AlcaldeTyagi2023a}, preferred observables and their star-polarized representations were constructed from Kähler geometry. Here the flat case is completed by proving that, among quadratic Hamiltonians, the preferred observables are exactly those whose star commutators reproduce their classical Poisson derivations. In~\cite{AlcaldeTyagi2023b}, statistical observables were obtained by smoothing with a probability density, and the relation between a prior and a choice of ordering was proposed. Here that relation is proved for Gaussian priors: the star-product equivalence operator is precisely Gaussian smoothing by the covariance cell. In~\cite{Alcalde2025}, coherent-state bundles over the Siegel space were proposed using the metaplectic and creation--annihilation formalisms. Here the admissible covariances are characterized by Williamson's theorem, and the corresponding Fock spaces and their transport are derived from classical amplitudes and Fronsdal's star polarization. Karabegov developed the systematic theory of star products with separation of variables on Kähler manifolds \cite{Karabegov1996}. The decomposition into a fixed antisymmetric part and a symmetric expression parameter, together with the equivalence operators between different expressions, is used here in the form developed by Omori, Maeda, Miyazaki, and Yoshioka \cite{OMMY2007}. The contribution made here is to identify the positive symmetric parameter with a classical covariance tensor. The Weyl--Wick heat equivalence is standard in phase-space analysis \cite{Folland1989}. Its interpretation here is statistical: the heat operator smooths observables by a classical Gaussian confidence density, and the associated Fock inner product integrates against that same density. \subsection{Conventions}\label{subsec:conventions} Throughout, \(\iota_{X_H}\omega=dH\) and \(\{F,G\}=\omega(X_F,X_G)\), so that \[ X_G(F)=\{F,G\}. \] For \(A\in\sp(V,\omega)\), the quadratic Hamiltonian is \begin{equation}\label{eq:HA} H_A(x) = -\frac12\omega(x,Ax) = \frac12\omega(Ax,x), \qquad X_{H_A}(x)=Ax, \qquad \{H_A,H_B\}=H_{[A,B]}. \end{equation} We write \[ \dim_{\mathbb R}V=2N, \qquad \mu_\omega=\frac{\omega^N}{N!}, \] and \(\lambda>0\) always denotes the covariance scale. \Cref{app:signs} lists the remaining sign conventions and \cref{tab:notation} summarizes the notation.

\begin{table}[ht]
\centering\small
\caption{Notation used throughout.}\label{tab:notation}
\begin{tabular}{@{}ll@{}}
\toprule
Symbol & Meaning \\
\midrule
$(P,\omega)$, $(V,\omega)$ & phase space; linear symplectic space \\
$\Mot$, $\omega_\Mot$ & space of motions; Souriau form \\
$\mathbb D=V^-\times V$, $\Omega_{\mathbb D}$ & real symplectic double \\
$\Dbl=\overline{V_J}\times V_J$ & complex Kähler double, coordinates $(\wb,z)$ \\
$M_f$ & multiplication operator by $f$ \\
$\Koop$ & Koopman generator \\
$g_J$, $G_J$; $\Sigma_J=G_J^{-1}$ & precision metric, its Hermitian matrix; covariance tensor \\
$\rho_{J,\lambda}$ & Gaussian cell density, covariance $\lambda\Sigma_J$ \\
$\tau$ & Siegel matrix, $\tau^{\T}=\tau$, $\operatorname{Im}\tau>0$ \\
$\Jp$ & space of compatible complex structures (Siegel space) \\
$\mathfrak k_J,\ \mathfrak p_J$ & Cartan splitting of $\sp$ at $J$ \\
$\Pol$, $\mathcal P_{J,\lambda}$ & polynomials on $\Dbl$; with the product $\star_{J,\lambda}$ \\
$T_{J,\lambda}$ & Gaussian smoothing operator of the cell \\
$\mathfrak K^\chi_{J,\lambda}$ & star-polarized sector with character $\chi$ \\
$\Fock$ & Fock space of the cell \\
\bottomrule
\end{tabular}
\end{table}

\section{Classical probability amplitudes and Koopman transport}\label{sec:amplitudes}

\paragraph{Motions and events.}
Let $(P,\omega)$ be a symplectic manifold, $H\in C^\infty(P)$, and $\Phi_t$ the flow of $X_H$.
The \emph{space of motions} $\Mot$ is the set of solutions $\gamma$ of $\dot\gamma=X_H(\gamma)$.
For each $t$, evaluation $\mathrm{ev}_t(\gamma)=\gamma(t)$ identifies $\Mot$ locally with $P$, and different evaluations differ by the flow, $\mathrm{ev}_t=\Phi_{t-t_0}\circ\mathrm{ev}_{t_0}$.
Because $\Phi_t$ is symplectic, $\omega_\Mot:=\mathrm{ev}_{t_0}^*\omega$ does not depend on $t_0$; it is Souriau's symplectic form on $\Mot$~\cite{Souriau1970}.
A point of $P$ is thus a \emph{representative} of a motion; the motion itself is the intrinsic object.

Fix a compact set $K$ of motions, invariant under the flow.
Its continuous observables form the commutative C$^*$-algebra $C(K)$, whose characters are the evaluations $f\mapsto f(m)$.
Deterministic motions are therefore the pure states, a general preparation is a probability measure $\rho$ on $K$ acting by $\omega_\rho(f)=\int_K f\,d\rho$, and the operational pairing between preparations and observations is the duality between measures and functions.

\paragraph{Amplitudes.}
Choose a reference measure $\mu$ and let $\rho\ll\mu$ with density $p_\rho$.
Its \emph{amplitude} is $\psi_\rho=\sqrt{p_\rho}\in L^2(K,\mu)$, and
\begin{equation}\label{eq:pairing}
\omega_\rho(f)=\langle\psi_\rho,M_f\psi_\rho\rangle,\qquad
\rho(\Delta)=\langle\psi_\rho,E(\Delta)\psi_\rho\rangle,\quad E(\Delta)=M_{\mathbf 1_\Delta}.
\end{equation}
The second identity has the form of the Born rule but is only $\rho(\Delta)=\int_\Delta p_\rho\,d\mu$ in Hilbert-space notation.
Two facts make the representation intrinsic.
First, replacing $\mu$ by $d\mu'=r\,d\mu$ replaces $\psi_\rho$ by $r^{-1/2}\psi_\rho$ and leaves $|\psi_\rho|^2d\mu$ unchanged, so the true object is the half-density $\sqrt{d\rho}$~\cite{Accardi1976}.
Second, if $|u|=1$ then $\langle u\psi,M_fu\psi\rangle=\langle\psi,M_f\psi\rangle$ for every $f$: pointwise phases are invisible to commuting observables, and $\sqrt{p_\rho}$ is the canonical positive representative of its class.

\begin{remark}[Multiplicity]
A general representation of $L^\infty(K,\mu)$ is a direct integral $\int^\oplus\mathcal H_m\,d\mu(m)$ with $(\pi(f)\Psi)(m)=f(m)\Psi(m)$; the induced density is $\|\Psi(m)\|^2$, and the commutant consists of the decomposable operators.
This paper uses only the multiplicity-free case $\mathcal H_m\simeq\C$.
\end{remark}

\paragraph{Koopman transport.}
If $\mu$ is invariant, $(U_t\psi)(m)=\psi(\Phi_{-t}m)$ is unitary and implements the flow on observables, $U_tM_fU_t^{-1}=M_{f\circ\Phi_{-t}}$.
If $\mu$ is only quasi-invariant, unitarity is restored by the half-density Jacobian,
\begin{equation}\label{eq:koopman}
(U_t\psi)(m)=\Bigl(\tfrac{d(\mu\circ\Phi_{-t})}{d\mu}(m)\Bigr)^{1/2}\psi(\Phi_{-t}m),
\end{equation}
and $U_t\psi_{\rho_0}$ is the amplitude of $\rho_t=(\Phi_t)_*\rho_0$ (compare~\cite{AlcaldeTyagi2023b}, where the same operator underlies the statistical model of signals).
Writing $U_t=e^{-it\Koop}$, the generator is $\Koop=-iX_H$ on a smooth core, with spectral resolution $\Koop=\int\nu\,dE_\Koop(\nu)$.
Its spectrum consists of the \emph{frequencies} of classical observables; it is not an energy spectrum.

\section{Linearization on one fibre}\label{sec:linear} The Koopman operator introduced in \cref{sec:amplitudes} describes the transport of amplitudes on the full space of motions. We now restrict attention to a preparation concentrated near a chosen reference motion \[ c(t)=\Phi_t(c(0)). \] Its first-order fluctuations are described by variations \(\delta c(t)\in T_{c(t)}P\). The relevant tangent space therefore changes with time. Before these variations can be compared, or represented by a single linear evolution equation, they must be transported to a common vector space. A symplectic connection provides precisely this identification. It separates two geometrically distinct operations: \begin{enumerate}[label=(\alph*)] \item parallel transport between the tangent spaces \(T_{c(t)}P\) along the reference motion; \item the internal linearized dynamics of the fluctuation once all variations have been expressed in one fixed tangent space. \end{enumerate} This distinction is important. It avoids identifying tangent spaces by means of a coordinate chart and makes the resulting linearization intrinsic. \paragraph{The fixed fibre.} An affine connection \(\nabla\) on the symplectic manifold \((P,\omega)\) is called \emph{symplectic} if \[ \nabla\omega=0. \] Equivalently, for vector fields \(X,Y,W\), \[ X\bigl[\omega(Y,W)\bigr] = \omega(\nabla_XY,W)+\omega(Y,\nabla_XW). \] Consequently, parallel transport along the reference motion, \[ \tau_{t,s}:T_{c(s)}P\longrightarrow T_{c(t)}P, \] preserves the symplectic form: \[ \omega_{c(t)}\bigl(\tau_{t,s}u,\tau_{t,s}w\bigr) = \omega_{c(s)}(u,w). \] Thus every \(\tau_{t,s}\) is a linear symplectic isomorphism. Fix the initial tangent space \[ V:=T_{c(0)}P, \qquad \omega_0:=\omega_{c(0)}. \] A variation \(\delta c(t)\in T_{c(t)}P\) is represented in this fixed space by parallel transport back to \(t=0\): \begin{equation}\label{eq:pullback} v(t):=\tau_{0,t}\,\delta c(t)\in V. \end{equation} The curve \(v(t)\) is therefore not a second physical trajectory. It is the first variation of the original trajectory, expressed at every time in the same symplectic vector space. Assume first that \(\nabla\) is torsion free. Since \(\delta c\) and \(\dot c\) arise from a two-parameter family of solutions, their covariant derivatives commute: \[ \nabla_{\dot c}\delta c = \nabla_{\delta c}\dot c = \nabla_{\delta c}X_H. \] Transporting this equation to \(V\) gives the fixed-fibre variational equation \begin{equation}\label{eq:variational} \dot v(t)=Z_t v(t), \qquad Z_t := \tau_{0,t}\circ(\nabla X_H)_{c(t)}\circ\tau_{t,0} \in\operatorname{End}(V). \end{equation} Thus \(Z_t\) is the linearization of the Hamiltonian vector field along \(c(t)\), expressed in the initial tangent space. If the connection has torsion \(\Theta\), the interchange of \(\dot c\) and \(\delta c\) produces the correction \[ \tau_{0,t}\circ \Theta(X_H,\cdot)_{c(t)} \circ\tau_{t,0}. \] The torsion-free assumption is therefore not required for the fixed-fibre construction itself; it is used below to identify the quadratic generator directly with the covariant Hessian of \(H\). \begin{proposition}[Linearized Koopman dynamics] \label{prop:linearKoopman} Let \(\nabla\) be a torsion-free symplectic connection. Then: \begin{enumerate}[label=(\roman*)] \item the generator \(Z_t\) is infinitesimally symplectic, \[ Z_t\in\mathfrak{sp}(V,\omega_0), \] and \eqref{eq:variational} is the Hamiltonian system generated by the time-dependent quadratic Hamiltonian \begin{equation}\label{eq:Ht} H_t(v) = \tfrac12\,\omega_0(Z_tv,v) = \tfrac12\,(\nabla^2H)_{c(t)} \bigl(\tau_{t,0}v,\tau_{t,0}v\bigr); \end{equation} \item if \(S(t)\) is the fundamental solution \[ \dot S(t)=Z_tS(t), \qquad S(0)=I, \] then \(S(t)\) is symplectic and the tangent flow factors as \[ T_{c(0)}\Phi_t=\tau_{t,0}S(t); \] \item the pullback \[ (U_ta)(v):=a\bigl(S(t)^{-1}v\bigr) \] defines a unitary propagator on \(L^2(V,\vol)\). It satisfies \[ i\partial_ta_t=\Koop_ta_t, \qquad \Koop_t=-iX_{H_t}=-i\{\,\cdot\,,H_t\}. \] \end{enumerate} \end{proposition} \begin{proof} We prove the three statements in turn. \smallskip \noindent \emph{Infinitesimal symplecticity.} Differentiate the Hamiltonian identity \[ \iota_{X_H}\omega=dH \] covariantly in the direction \(u\). Since \(\nabla\omega=0\), \[ \omega(\nabla_uX_H,w) = (\nabla^2H)(u,w). \] For a torsion-free connection, the covariant Hessian is symmetric. Hence \[ \omega(\nabla_uX_H,w) = \omega(\nabla_wX_H,u) = -\omega(u,\nabla_wX_H). \] Thus \(\nabla X_H\) is infinitesimally symplectic. Because parallel transport is symplectic, its conjugate \(Z_t\) satisfies \[ \omega_0(Z_tu,w)+\omega_0(u,Z_tw)=0, \] and therefore \(Z_t\in\mathfrak{sp}(V,\omega_0)\). For a linear vector field \(v\mapsto Zv\) with \(Z\in\mathfrak{sp}(V,\omega_0)\), the quadratic function \[ H_Z(v)=\tfrac12\,\omega_0(Zv,v) \] satisfies \(\iota_{X_{H_Z}}\omega_0=dH_Z\) and has Hamiltonian vector field \(X_{H_Z}(v)=Zv\). Applying this observation to \(Z_t\) gives the first expression in \eqref{eq:Ht}. The second follows from \[ \omega(\nabla_uX_H,w)=(\nabla^2H)(u,w) \] and the symplecticity of \(\tau_{t,0}\). \smallskip \noindent \emph{Symplectic fundamental solution.} For \(u,w\in V\), \begin{align*} \frac{d}{dt}\, \omega_0\bigl(S(t)u,S(t)w\bigr) &= \omega_0\bigl(Z_tS(t)u,S(t)w\bigr) \\ &\quad+ \omega_0\bigl(S(t)u,Z_tS(t)w\bigr) =0. \end{align*} Since \(S(0)=I\), it follows that \[ \omega_0\bigl(S(t)u,S(t)w\bigr)=\omega_0(u,w). \] Thus \(S(t)\) is symplectic. The factorization \[ T_{c(0)}\Phi_t=\tau_{t,0}S(t) \] states that the tangent flow is obtained by first evolving the fluctuation internally in \(V\), through \(S(t)\), and then transporting the result from \(T_{c(0)}P\) to \(T_{c(t)}P\). Both sides satisfy the same variational equation along \(c\) and have the same initial value, so they coincide. \smallskip \noindent \emph{Koopman transport.} A symplectic linear map preserves the Liouville volume \(\vol\). Consequently, \[ \int_V\left|a\bigl(S(t)^{-1}v\bigr)\right|^2\vol(v) = \int_V|a(v)|^2\vol(v), \] and \(U_t\) is unitary. Finally, differentiating \[ a_t(v)=a\bigl(S(t)^{-1}v\bigr) \] gives \[ \partial_ta_t=-X_{H_t}a_t. \] Multiplication by \(i\) yields \[ i\partial_ta_t = -iX_{H_t}a_t = -i\{a_t,H_t\} = \Koop_ta_t. \] \end{proof} The proposition gives a geometric factorization of the tangent dynamics: \[ \underbrace{T_{c(0)}P}_{\text{initial fibre}} \xrightarrow{\;S(t)\;} \underbrace{T_{c(0)}P}_{\text{internal evolution}} \xrightarrow{\;\tau_{t,0}\;} \underbrace{T_{c(t)}P}_{\text{current fibre}}. \] The connection handles the comparison of different tangent spaces, whereas \(S(t)\) contains the actual linearized Hamiltonian dynamics. This also explains why the ordinary coordinate Hessian of \(H\) is not the correct intrinsic generator along a non-critical trajectory. Second coordinate derivatives do not transform tensorially away from a critical point. The covariant Hessian \(\nabla^2H\), together with parallel transport, supplies exactly the correction needed to define the quadratic Hamiltonian \(H_t\) intrinsically. \paragraph{Gaussian amplitudes on the fibre.} Let a statistical preparation concentrated near \(c(0)\) be represented on \(V\) by a probability density \(p_0\) with respect to the Liouville volume \(\vol\). Its canonical positive amplitude is \[ \psi_0=\sqrt{p_0}. \] Under the fixed-fibre dynamics, both the density and its square-root amplitude are transported by pullback: \[ p_t(v)=p_0\bigl(S(t)^{-1}v\bigr), \qquad \psi_t(v)=\psi_0\bigl(S(t)^{-1}v\bigr). \] No Jacobian factor occurs because \(S(t)\) is symplectic and therefore preserves \(\vol\). In particular, \[ \int_V p_t\,\vol=1, \qquad \|\psi_t\|_{L^2(V,\vol)}=1. \] For every observable \(a\) on the fluctuation space, \[ \int_V a(v)p_t(v)\,\vol(v) = \langle\psi_t,M_a\psi_t\rangle. \] This is the classical preparation--observation pairing written in square-root form. Suppose now that \(p_0\) is a centred Gaussian with covariance \(\Sigma_0\) and precision \[ G_0:=\Sigma_0^{-1}. \] Since \(\dim V=2N\), the transported Gaussian amplitude is \begin{equation}\label{eq:gauss-fibre} \psi_t(v) = \frac{1} {(2\pi)^{N/2}(\det\Sigma_t)^{1/4}} \exp\left(-\tfrac14\,v^{\T}G_tv\right), \qquad \Sigma_t=S(t)\Sigma_0S(t)^{\T}, \qquad G_t=S(t)^{-\T}G_0S(t)^{-1}. \end{equation} These two transformation laws express the dual roles of covariance and precision: \begin{itemize} \item the covariance \(\Sigma_t\), being a positive contravariant two-tensor, is pushed forward by the linearized motion; \item the precision \(G_t=\Sigma_t^{-1}\), being a positive covariant two-tensor, is pulled back by the inverse motion. \end{itemize} Differentiating these expressions and using \(\dot S(t)=Z_tS(t)\) gives \[ \dot\Sigma_t = Z_t\Sigma_t+\Sigma_tZ_t^{\T}, \qquad \dot G_t = -Z_t^{\T}G_t-G_tZ_t. \] The first equation follows the evolving width and orientation of the confidence cell. The second follows the quadratic form appearing in the Gaussian exponent. Because every symplectic matrix has determinant one, \[ \det\Sigma_t = (\det S(t))^2\det\Sigma_0 = \det\Sigma_0. \] The covariance cell may therefore stretch in some directions, contract in others, and undergo shear, while preserving its symplectic volume. Correspondingly, the amplitude changes shape but retains unit \(L^2\)-norm. This deformation is ordinary classical Liouville transport. It does not require, and should not be interpreted as, interference. \paragraph{Summary.} The construction developed so far can be organized into three successive levels: \[ \begin{aligned} &(P,\omega,H) \longrightarrow (\Mot,\omega_{\Mot}) \longrightarrow C(K), \\[2mm] &\rho \longmapsto \sqrt{d\rho}, \qquad \Phi_t \longmapsto U_t, \\[2mm] &T_{c(0)}\Phi_t = \tau_{t,0}S(t), \qquad a \longmapsto a\circ S(t)^{-1}. \end{aligned} \] The first line gives the classical event algebra on the space of motions. The second realizes preparations by square-root probability amplitudes and transports them by the nonlinear Koopman operator. The third is the intrinsic linearization of this transport near one reference motion: the symplectic connection compares tangent spaces, while the quadratic Hamiltonian \(H_t\) generates the internal evolution of fluctuations in a fixed fibre. From now on we work in this fixed fibre and write \[ (V,\omega):=(T_{c(0)}P,\omega_0). \]
\section{Covariance cells as positive polarizations}\label{sec:cells} The preceding section described a localized preparation by a Gaussian amplitude on the fixed symplectic fibre \((V,\omega)\). Its covariance tensor specifies the size, shape, and orientation of the corresponding confidence region. A general positive covariance, however, contains more data than a compatible complex structure: in Williamson normal form it carries one positive scale for each symplectic two-plane. The present paper singles out the covariances for which these intrinsic symplectic scales coincide. We call them \emph{balanced covariance cells}. The adjective ``balanced'' refers to the symplectic geometry, not to the appearance of the cell in a particular coordinate system. A balanced cell need not be a Euclidean ball. It may be stretched, tilted, and sheared, but it is obtained from a standard isotropic cell by a symplectic transformation. Equivalently, it assigns the same covariance scale to every conjugate symplectic plane. This restriction has a geometric purpose. Once the common scale is separated out, the remaining shape of the cell is described exactly by a compatible complex structure \(J\), or equivalently by a positive Lagrangian polarization. Thus a balanced covariance cell is not merely a Gaussian ellipsoid placed on a symplectic vector space: its covariance geometry, Kähler metric, and positive polarization are different descriptions of the same object. We shall establish the correspondence \[ \boxed{ \text{balanced covariance cell} \ \longleftrightarrow\ \text{compatible complex structure} \ \longleftrightarrow\ \text{positive Lagrangian polarization} } \] and then describe how quadratic Hamiltonian flows transport these objects. The common covariance scale will be denoted by \(\lambda>0\); the normalized geometric shape will be encoded by \(J\). \subsection{Four descriptions of one object} A complex structure \(J\) on \(V\) is \emph{compatible} with \(\omega\) if \[ J^2=-I, \qquad \omega(Ju,Jv)=\omega(u,v), \] and the bilinear form \[ g_J(u,v):=\omega(u,Jv) \] is positive definite. The three conditions have distinct roles: \(J^2=-I\) makes \(V\) into a complex vector space, preservation of \(\omega\) makes \(J\) symplectic, and positivity turns the pair \((\omega,J)\) into a linear Kähler structure. The complexification \(V_\C\) then splits into the eigenspaces of \(J\): \[ V_\C=L_J\oplus\overline{L_J}, \qquad L_J:=\ker(J-iI). \] The subspace \(L_J\) is complex Lagrangian, and it is positive in the sense that \[ h_J(\xi,\eta) := -i\,\omega_\C(\xi,\bar\eta) \] defines a positive Hermitian form on \(L_J\). Indeed, if \(\xi=u-iJu\), then \[ h_J(\xi,\xi)=2g_J(u,u)>0 \] for \(u\neq0\). Conversely, suppose that \(L\subset V_\C\) is a positive complex Lagrangian satisfying \[ V_\C=L\oplus\bar L. \] There is then a unique compatible complex structure determined by \[ J|_L=iI, \qquad J|_{\bar L}=-iI. \] Compatible complex structures and positive Lagrangian subspaces are therefore equivalent descriptions of the same positive symplectic geometry. The metric \(g_J\) measures displacement relative to this geometry. We call it the \emph{precision metric}, because it appears in the exponent of the associated Gaussian. Its inverse \[ \Sigma_J:=g_J^{-1} \] is the corresponding \emph{covariance tensor}. Thus \[ g_J \quad\text{acts on vectors, whereas}\quad \Sigma_J \quad\text{acts on covectors}. \] In a linear frame they are represented by inverse positive matrices, but intrinsically they have opposite tensor character. \paragraph{Why a balance condition is needed.} Not every positive covariance tensor is of the form \(\lambda\Sigma_J\). To see what distinguishes those that are, recall that Williamson's theorem associates with every positive covariance tensor \(C\) a collection of positive numbers \[ \nu_1(C),\ldots,\nu_N(C), \] its \emph{symplectic eigenvalues}. In a suitable symplectic frame, \[ C = S \begin{pmatrix} D&0\\ 0&D \end{pmatrix} S^{\T}, \qquad D=\operatorname{diag}(\nu_1,\ldots,\nu_N), \qquad S\in\Sp(V,\omega). \] The numbers \(\nu_j\) are invariant under symplectic changes of frame. They are therefore the intrinsic covariance scales carried by the individual symplectic planes. Ordinary Euclidean eigenvalues depend on the selected coordinates; Williamson eigenvalues do not. A covariance is \emph{balanced} when all these intrinsic scales agree: \[ \nu_1=\cdots=\nu_N=\lambda. \] After removing the common scale \(\lambda\), no independent dilation remains in one symplectic plane relative to another. What remains is precisely the shape and orientation data carried by a compatible complex structure. \begin{proposition}[Balanced cells]\label{prop:balanced}

Let \(\Sigma\in\operatorname{Sym}^2(V)\) be positive definite, regarded equivalently as a positive symmetric bilinear form on \(V^*\) and let \(\lambda>0\). Then \[ \Sigma=\lambda\Sigma_J \] for a compatible complex structure \(J\) if and only if all the symplectic eigenvalues of \(\Sigma\) are equal to \(\lambda\). When such a \(J\) exists, it is unique. Consequently, \[ J \ \longleftrightarrow\ L_J \ \longleftrightarrow\ g_J \ \longleftrightarrow\ \Sigma_J \] are bijective descriptions of, respectively, \begin{enumerate}[label=(\roman*)] \item compatible complex structures; \item positive Lagrangian subspaces of \(V_\C\); \item positive symplectic metrics on \(V\); \item covariance tensors whose symplectic eigenvalues are all equal to one. \end{enumerate} \end{proposition} \begin{proof} Choose a Darboux frame and write \[ \omega(u,v)=u^{\T}\Omega v, \qquad \Omega^{\T}=-\Omega, \qquad \Omega^2=-I. \] For a compatible complex structure \(J\), the precision metric is represented by \[ g_J=\Omega J. \] Compatibility implies that \(g_J\) is symmetric and positive definite. Moreover, \[ g_J^{\T}\Omega g_J = J^{\T}\Omega J = \Omega, \] so \(g_J\) is itself symplectic. Conversely, let \(g\) be a positive symmetric symplectic matrix and set \[ J:=\Omega^{-1}g. \] The symplectic identity \[ g^{\T}\Omega g=\Omega \] reduces, since \(g^{\T}=g\), to \(g\Omega g=\Omega\). It follows that \(J^2=-I\), while \(g=\Omega J\) gives \[ g(u,v)=\omega(u,Jv). \] Thus \(J\) is compatible with \(\omega\). The relation \(g=\Omega J\) also proves uniqueness. It remains to characterize the positive symplectic matrices. By Williamson's theorem~\cite{Williamson1936}, every positive symmetric matrix can be written as \[ g=S^{\T} \begin{pmatrix} D&0\\ 0&D \end{pmatrix} S, \qquad S\in\Sp(V,\omega), \] where \(D=\operatorname{diag}(\nu_1,\ldots,\nu_N)\). Since \(S\) is symplectic, \(g\) is symplectic if and only if its Williamson diagonal is symplectic. This occurs precisely when \[ \nu_1=\cdots=\nu_N=1. \] Finally, if \(\Sigma=\lambda\Sigma_J\), then \[ g_J = \Sigma_J^{-1} = \lambda\Sigma^{-1}. \] Hence \(g_J\) is positive and symplectic exactly when \(\lambda\Sigma^{-1}\) has unit symplectic eigenvalues, or equivalently when all symplectic eigenvalues of \(\Sigma\) equal \(\lambda\). \end{proof} \begin{definition}[Balanced covariance cell] Let \(J\) be compatible with \(\omega\) and let \(\lambda>0\). The \emph{balanced covariance cell} associated with \((J,\lambda)\) is the ellipsoid \[ \mathcal C_{J,\lambda} := \left\{ x\in V: g_J(x,x)\leq\lambda \right\}. \] Equivalently, it is the centred Gaussian confidence cell with covariance tensor \[ \lambda\Sigma_J=\lambda g_J^{-1}. \] \end{definition} The parameter \(\lambda\) determines the common symplectic scale of the cell. The complex structure \(J\) determines its normalized shape and orientation. This separation is useful because symplectic transport changes \(J\), and hence deforms the cell, while preserving its Williamson spectrum. General Gaussian covariances need not be balanced. If their Williamson eigenvalues are unequal, each symplectic plane carries an independent scale: \[ \lambda_1,\ldots,\lambda_N. \] Such covariances contain both polarization data and additional plane-dependent scale data. The balanced class is the homogeneous subfamily in which the positive polarization alone determines the normalized covariance geometry. We return to the general case in \cref{sec:outlook}. \subsection{The space of cells and its Cartan splitting} Fixing the common scale \(\lambda\), the remaining freedom in a balanced cell is precisely the choice of compatible complex structure. The symplectic group acts transitively on these structures by \[ S\cdot J:=SJS^{-1}. \] If \(J_0\) is a reference complex structure, its stabilizer consists of those symplectic transformations that commute with \(J_0\): \[ \mathrm U(V,\omega,J_0) := \left\{ S\in\Sp(V,\omega): SJ_0=J_0S \right\}. \] These transformations preserve \(\omega\), \(J_0\), \(g_{J_0}\), and \(\Sigma_{J_0}\) simultaneously. They therefore preserve the reference cell. It follows that the space of balanced cells at fixed scale is the Hermitian symmetric space \[ \Jp \simeq \Sp(V,\omega)/\mathrm U(V,\omega,J_0). \] This space was proposed in~\cite{Alcalde2025} as the statistical manifold of linearized Hamiltonian preparations. After choosing a symplectic frame, \(\Jp\) may be realized as the Siegel upper half space \[ \mathfrak H_N = \left\{ \tau\in\operatorname{Mat}_{N\times N}(\C): \tau^{\T}=\tau,\ \operatorname{Im}\tau>0 \right\} \] \cite{Siegel1943}. In this realization, the positive Lagrangian subspace is the graph \[ L_J = \left\{ (q,p)\in V_\C: p=\tau q \right\}. \] The Siegel matrix \(\tau\) is complex symmetric and parametrizes the positive Lagrangian graph. It should not be confused with the Hermitian matrix \(G_J\) representing the precision metric below. The two objects encode the same positive geometry in different realizations, but they do not have the same tensorial role. At the infinitesimal level, \(J\) defines the Cartan involution \[ \vartheta_J(A):=JAJ^{-1} \] on \(\sp(V,\omega)\). Its eigenspace decomposition is \begin{equation}\label{eq:cartan} \sp(V,\omega) = \mathfrak k_J\oplus\mathfrak p_J, \qquad \mathfrak k_J = \{A\in\sp:[A,J]=0\} \simeq\mathfrak u(N), \qquad \mathfrak p_J = \{A\in\sp:AJ+JA=0\}. \end{equation} The associated projections are \[ A_{\mathfrak k} = \tfrac12(A-JAJ), \qquad A_{\mathfrak p} = \tfrac12(A+JAJ). \] They satisfy the symmetric-space relations \[ [\mathfrak k,\mathfrak k]\subseteq\mathfrak k, \qquad [\mathfrak k,\mathfrak p]\subseteq\mathfrak p, \qquad [\mathfrak p,\mathfrak p]\subseteq\mathfrak k. \] The geometric meaning of this decomposition is immediate: \begin{itemize} \item \(\mathfrak k_J\) contains the infinitesimal transformations that preserve the cell; \item \(\mathfrak p_J\) contains the infinitesimal transformations that move the cell through the space \(\Jp\). \end{itemize} After complexification, \[ \mathfrak p_J^\C = \mathfrak p_J^+\oplus\mathfrak p_J^-, \] where \(\mathfrak p_J^+\) and \(\mathfrak p_J^-\) are, respectively, the holomorphic and antiholomorphic tangent spaces of \(\Jp\) at \(J\). \paragraph{Transport of cells.} Let \(H_A\) be a quadratic Hamiltonian whose linear Hamiltonian vector field is generated by \(A\in\sp(V,\omega)\), and let \[ S_t=e^{tA}. \] The flow transports every description in \cref{prop:balanced} simultaneously: \begin{equation}\label{eq:transport} J_t=S_tJ_0S_t^{-1}, \qquad L_{J_t}=S_tL_{J_0}, \qquad g_t=S_t^{-\T}g_0S_t^{-1}, \qquad \Sigma_t=S_t\Sigma_0S_t^{\T}. \end{equation} The first two expressions describe transport of the complex structure and its positive polarization. The last two are the dual transformation laws for precision and covariance. Differentiating at time \(t\) gives \[ \dot J_t=[A,J_t], \qquad \dot\Sigma_t = A\Sigma_t+\Sigma_tA^{\T}. \] Because symplectic congruence preserves Williamson eigenvalues, a balanced covariance remains balanced and retains the same scale \(\lambda\). The flow changes its shape and orientation, not its intrinsic symplectic spectrum. \begin{proposition}[Rotations and squeezes]\label{prop:rotsqueeze} Let \(J\) be compatible with \(\omega\). Then \[ A\in\mathfrak k_J \quad\Longleftrightarrow\quad A\Sigma_J+\Sigma_JA^{\T}=0. \] Equivalently, \(A\in\mathfrak k_J\) if and only if the flow generated by \(H_A\) fixes the covariance cell. Moreover, the map \[ \mathfrak p_J \longrightarrow T_{\Sigma_J}\Jp, \qquad A \longmapsto A\Sigma_J+\Sigma_JA^{\T}, \] is an isomorphism. \end{proposition} \begin{proof} The stabilizer of \(J\) in \(\Sp(V,\omega)\) is the unitary group \(\mathrm U(V,\omega,J)\). Since \[ g_J=\omega(\,\cdot\,,J\,\cdot\,), \qquad \Sigma_J=g_J^{-1}, \] a symplectic transformation preserves \(J\) if and only if it preserves \(g_J\), and hence if and only if it preserves \(\Sigma_J\). Differentiating the congruence action \[ \Sigma\longmapsto S\Sigma S^{\T} \] at the identity gives \[ \delta_A\Sigma_J = A\Sigma_J+\Sigma_JA^{\T}. \] Its kernel is therefore precisely \(\mathfrak k_J\). Since \[ \sp(V,\omega) = \mathfrak k_J\oplus\mathfrak p_J, \] the restriction of the infinitesimal orbit map to \(\mathfrak p_J\) is injective and identifies \(\mathfrak p_J\) with the tangent space of the homogeneous orbit \(\operatorname{Sp}(V,\omega)/\mathrm U(V,\omega,J)\) at \(\Sigma_J\).\end{proof} The terminology ``rotations and squeezes'' refers to these two infinitesimal roles. Elements of \(\mathfrak k_J\) may rotate points inside the cell, but they leave the covariance tensor itself fixed. Elements of \(\mathfrak p_J\) squeeze, shear, or reorient the cell and therefore move it through \(\Jp\). A general quadratic Hamiltonian contains both parts: \[ A=A_{\mathfrak k}+A_{\mathfrak p}. \] Its \(\mathfrak k\)-component acts within the selected resolution geometry, whereas its \(\mathfrak p\)-component transports that geometry. Quadratic Hamiltonian dynamics can therefore be read as the symplectic transport of the resolution structure itself. \begin{example}[\(N=1\)]\label{ex:N1} Let \(V=\R^2\) with coordinates \((q,p)\) and \(\omega=dq\wedge dp\). A balanced covariance cell of scale \(\lambda\) is an ellipse of area \(\pi\lambda\). All such ellipses are symplectic images of the circle \[ q^2+p^2\leq\lambda. \] Their space is \[ \Sp(2,\R)/\mathrm U(1) \simeq \mathrm{SL}(2,\R)/\mathrm{SO}(2), \] which is the upper half plane. For the standard complex structure, the compact direction \(\mathfrak k_J\) is generated by the rotational Hamiltonian \[ \tfrac12(q^2+p^2). \] It moves points around the circular cell without changing the cell. The noncompact space \(\mathfrak p_J\) is spanned by \[ \tfrac12(q^2-p^2), \qquad qp. \] These generators deform the circle into ellipses of the same area, with different eccentricities and orientations. \end{example} \subsection{The doubled spaces} Two doubled spaces appear in the paper. Although both combine two copies of the original linear space, they answer different questions: \begin{itemize} \item the \emph{real symplectic double} represents dynamical transport as a relation between initial and final configurations; \item the \emph{complex Kähler double} represents the comparison of configurations relative to a selected covariance cell. \end{itemize} Keeping these roles separate prevents the two constructions from being mistaken for alternative versions of the same doubling. \paragraph{The real double: transport as a relation.} Consider \[ \mathbb D:=V^-\times V, \qquad \Omega_{\mathbb D} := -\pi_1^*\omega+\pi_2^*\omega, \] where \(V^-\) denotes \(V\) equipped with the symplectic form \(-\omega\). The graph of the linear symplectic flow, \[ \Gamma_t := \left\{ (x,S_tx): x\in V \right\}, \] is Lagrangian in \(\mathbb D\). Indeed, for the graph embedding \(\iota_t(x)=(x,S_tx)\), \[ \iota_t^*\Omega_{\mathbb D} = -\omega+S_t^*\omega = 0. \] At \(t=0\), the graph is the diagonal: \[ \Gamma_0 = \left\{ (x,x): x\in V \right\}. \] Hamiltonian transport may therefore be viewed as a deformation of the diagonal through Lagrangian correspondences \cite{GuilleminSternberg1984}. At the same time, the flow transports the positive Lagrangian polarization in the complexified single space: \[ L_{J_0}\subset V_\C \quad\longmapsto\quad L_{J_t}=S_tL_{J_0}. \] These two Lagrangian objects encode complementary information. The real graph \(\Gamma_t\) records \emph{which} configurations are related by the dynamics. The positive Lagrangian \(L_{J_t}\) records the \emph{resolution geometry} through which those configurations are described. \paragraph{The complex double: comparison of cells.} Write \(V_J\) for the real vector space \(V\) regarded as a complex vector space through \(J\). The \emph{Kähler double} is \[ \Dbl:=\overline{V_J}\times V_J. \] Its coordinates \((\wb,z)\) are independent. The original real space is embedded as the real diagonal \[ \wb=\bar z. \] If \(z\) and \(w\) are the centres of two cells with the same covariance geometry, their separation relative to that geometry is \begin{equation}\label{eq:separation} D_J(z,w) = (z-w)^\dagger G_J(z-w). \end{equation} Thus \(D_J\) is not simply a coordinate distance. It measures the displacement between the two centres in units set by the cell's precision metric. A displacement along a direction of high precision is weighted more strongly than the same coordinate displacement along a direction of low precision. Equation \eqref{eq:separation} is the flat quadratic case of Calabi's diastasis~\cite{Calabi1953}. In \cref{sec:outlook}, this comparison function will be recovered from the overlap of two covariance cells. \subsection{Three faces of the covariance tensor} The covariance tensor introduced geometrically above also appears in two familiar analytic constructions. The same pair \[ G=\Sigma^{-1} \] governs: \begin{enumerate}[label=(\roman*)] \item the duality between velocity and momentum; \item the localization of a Gaussian confidence measure; \item the characteristic function of that measure. \end{enumerate} The next proposition records these three realizations and their common transport law. \begin{proposition}[Mechanical, probabilistic, and Fourier forms] \label{prop:threefaces} Let \[ \Sigma=\Sigma_J, \qquad G=G_J=\Sigma^{-1}. \] Then: \begin{enumerate}[label=(\roman*)] \item \emph{Legendre form.} The kinetic Lagrangian \[ \mathcal L(\dot x) = \tfrac12\langle G\dot x,\dot x\rangle \] has momentum \[ p=G\dot x \] and Hamiltonian \[ \mathcal H(p) = \tfrac12\langle p,\Sigma p\rangle. \] Because \(g_J=\omega(\,\cdot\,,J\,\cdot\,)\), the Legendre map may be written intrinsically as \[ p=-\omega^\flat J\dot x. \] \item \emph{Probability form.} The centred Gaussian measure associated with the cell has density \[ d\mu_\Sigma(x) \propto \exp\left( -\tfrac12\langle Gx,x\rangle \right)dx. \] \item \emph{Fourier form.} The characteristic function of this measure is \[ \chi_\Sigma(\xi) = \exp\left( -\tfrac12\langle\xi,\Sigma\xi\rangle \right). \] It is continuous, normalized, and of positive type. By Bochner's theorem, it determines the Gaussian measure \(\mu_\Sigma\) uniquely. \end{enumerate} Under the symplectic transport \eqref{eq:transport}, the three realizations transform coherently: \[ \mu_{\Sigma_t} = (S_t)_*\mu_{\Sigma_0}, \] \[ \sqrt{d\mu_{\Sigma_t}} = \sqrt{d\mu_{\Sigma_0}}\circ S_t^{-1}, \] and \[ \chi_{\Sigma_t}(\xi) = \chi_{\Sigma_0}(S_t^{\T}\xi). \] \end{proposition} \begin{proof} The Legendre transformation follows directly from \[ p = \frac{\partial\mathcal L}{\partial\dot x} = G\dot x, \qquad \dot x=\Sigma p. \] Therefore \[ \mathcal H(p) = \langle p,\dot x\rangle-\mathcal L(\dot x) = \tfrac12\langle p,\Sigma p\rangle. \] Since \[ g_J(u,v)=\omega(u,Jv), \] the metric isomorphism is \[ g_J^\flat=-\omega^\flat J, \] which gives the intrinsic expression for the momentum. The Gaussian density and its characteristic function are related by the standard Fourier transform of a positive quadratic Gaussian. The positive-type property follows either directly from the characteristic function representation or from Bochner's theorem. Finally, if \[ \Sigma_t=S_t\Sigma_0S_t^{\T}, \qquad G_t=S_t^{-\T}G_0S_t^{-1}, \] then \[ \langle G_tx,x\rangle = \langle G_0S_t^{-1}x,S_t^{-1}x\rangle. \] Since \(S_t\) preserves Liouville volume, this gives \[ \mu_{\Sigma_t}=(S_t)_*\mu_{\Sigma_0} \] and the stated Koopman transport of the square-root amplitude. Likewise, \[ \langle\xi,\Sigma_t\xi\rangle = \langle S_t^{\T}\xi,\Sigma_0S_t^{\T}\xi\rangle, \] which proves the transformation law for \(\chi_{\Sigma_t}\). \end{proof} The roles of the two inverse tensors can now be summarized succinctly: \[ \begin{array}{c|c} \text{Precision }G & \text{Covariance }\Sigma \\ \hline \text{quadratic Lagrangian} & \text{quadratic Hamiltonian} \\ \text{Gaussian density} & \text{characteristic function} \end{array} \] The Legendre transform exchanges precision and covariance between velocity and momentum variables. The Fourier transform exchanges them between a Gaussian density and its positive-type characteristic function. The Koopman operator transports both descriptions by the same symplectic flow. Normalization constants and the corresponding symplectic Fourier transform are collected in \cref{app:gauss}.
\section{The star product of a cell}\label{sec:star} A balanced covariance cell determines more than a Gaussian preparation. Its compatible complex structure \(J\) distinguishes holomorphic from antiholomorphic directions, while its covariance tensor \(\Sigma_J\) specifies how derivatives in those complementary directions are contracted. Together, these data define an associative algebra of observables adapted to the cell. The resulting product combines two geometrically distinct structures. Its antisymmetric part is fixed by the symplectic form \(\omega\) and reproduces the Poisson bracket. Its symmetric part is fixed by \(\Sigma_J\) and records the statistical resolution selected by the cell. Thus the symplectic structure determines the underlying noncommutative algebra, whereas the covariance cell determines the particular ordering, or expression, in which that algebra is realized. The complex polarization supplies a natural orientation to the contractions: antiholomorphic derivatives act on the left factor and holomorphic derivatives on the right. Consequently, holomorphic functions multiply undeformed from the left, and antiholomorphic functions multiply undeformed from the right. This is the \emph{separation-of-variables} property. \paragraph{Historical note.} Products of the separated form used below were obtained independently by the author in 1991 in a manuscript accepted for publication in the \emph{Journal of Mathematical Physics}, but ultimately unpublished. A later published presentation of the underlying Kähler construction appears in~\cite[Example~1]{AlcaldeTyagi2023a}.
We use  Karabegov's systematic theory of star products with separation of variables \cite{Karabegov1996}.The decomposition of a constant-coefficient star product into a fixed antisymmetric part and a variable symmetric ordering part, together with the differential equivalences between different orderings, is used in the form developed by Yoshioka and collaborators \cite{OMMY2007}. In the present setting, the symmetric ordering parameter is no longer arbitrary: it is identified with the positive covariance tensor of the cell. This identification is what allows the ordering transformation to be interpreted as Gaussian smoothing. 

\subsection{Coordinates on the Kähler double} Let \(z=(z^1,\ldots,z^N)\) be complex-linear coordinates on \(V_J\), characterized by \[ dz^a\circ J=i\,dz^a, \] and let \(\wb=(\wb^{\bar 1},\ldots,\wb^{\bar N})\) denote the independent conjugate coordinates on \(\overline{V_J}\). The Kähler double introduced in the preceding section is \[ \Dbl=\overline{V_J}\times V_J, \] with independent coordinates \((\wb,z)\). The original real vector space is recovered as the real form \[ \wb=\bar z. \] Let \[ G_J=(g_{a\bar b}) \] be the positive Hermitian matrix of the precision metric. With the normalization used throughout this section, the symplectic form on the real locus is \[ \omega = i\,g_{a\bar b}\, dz^a\wedge d\bar z^{\bar b}. \] Denote by \[ \Sigma_J^{\bar ba} \] the inverse Hermitian matrix, so that \[ \Sigma_J^{\bar ba}g_{a\bar c} = \delta^{\bar b}_{\bar c}. \] The polarized extension of the quadratic Kähler potential to the double is \begin{equation}\label{eq:potential} \Phi_J(\wb,z) := g_{a\bar b}\,z^a\wb^{\bar b}. \end{equation} Its mixed Hessian recovers the precision metric: \[ \partial_a\partial_{\bar b}\Phi_J = g_{a\bar b}. \] On the real form, \[ \phi_J(z,\bar z) := \Phi_J(\bar z,z) = \bar z^{\T}G_Jz. \] Thus \(\phi_J\) is the positive quadratic form appearing in the Gaussian density of the cell. The corresponding Poisson bracket on the double is \begin{equation}\label{eq:poisson-double} \{F,G\}_J = -i\,\Sigma_J^{\bar ba} \left( \partial_aF\,\partial_{\bar b}G - \partial_{\bar b}F\,\partial_aG \right), \end{equation} where \[ \partial_a:=\frac{\partial}{\partial z^a}, \qquad \partial_{\bar b}:= \frac{\partial}{\partial\wb^{\bar b}}. \] Equations \eqref{eq:potential} and \eqref{eq:poisson-double} exhibit the duality between precision and covariance. The mixed Hessian of the potential is \(G_J\), while its inverse \(\Sigma_J\) performs the Poisson contraction. The same inverse matrix will also govern the symmetric contraction in the star product. The covariance tensor therefore enters twice: antisymmetrically, through the Poisson bracket, and symmetrically, through the resolution selected by the cell. For \(N=1\), take \(G_J=1\) and \[ z=\frac{q+ip}{\sqrt2}, \qquad \bar z=\frac{q-ip}{\sqrt2}. \] Then \[ \phi_J=z\bar z=\frac12(q^2+p^2), \] and \eqref{eq:poisson-double} gives \[ \{z,\bar z\}=-i, \] consistently with \(\omega=dq\wedge dp\). We work initially with the polynomial algebra \[ \Pol := \C[ z^1,\ldots,z^N, \wb^{\bar1},\ldots,\wb^{\bar N} ]. \] It contains the holomorphic and antiholomorphic subalgebras \[ \mathcal O_J:=\C[z^1,\ldots,z^N], \qquad \overline{\mathcal O}_J := \C[\wb^{\bar1},\ldots,\wb^{\bar N}]. \] Restricting first to polynomials avoids convergence questions: every bidifferential expansion below terminates after finitely many terms.

\subsection{The separated product} The star product used here is the covariance-cell version of the flat Kähler product constructed in~\cite[Example~1]{AlcaldeTyagi2023a}. Its origin is the decomposition of Hamiltonian vector fields into holomorphic and antiholomorphic parts. For a preferred observable \(a\), the earlier construction expresses left and right star multiplication through, respectively, the holomorphic and antiholomorphic components of its Hamiltonian vector field. In the flat Kähler case, that prescription extends to the closed associative product below. For the covariance cell determined by \(J\), the inverse Hermitian metric \[ \Sigma_J=G_J^{-1} \] performs the contraction between the antiholomorphic derivative of the first factor and the holomorphic derivative of the second. Thus the positive polarization does not merely select a class of functions: it determines the direction of the deformation itself. \begin{definition}[The star product of a cell] \label{def:cell-star} Let \(J\) be compatible with \(\omega\), let \(\lambda>0\), and let \(F,G\in\Pol\). Define \begin{equation}\label{eq:star} \begin{split} F\st G &:= m\circ \exp\left( \lambda\, \Sigma_J^{\bar ab} \partial_{\bar a}\otimes\partial_b \right)(F\otimes G) \\[1mm] &= \sum_{r\geq0} \frac{\lambda^r}{r!}\, \Sigma_J^{\bar a_1b_1}\cdots \Sigma_J^{\bar a_rb_r} \left( \partial_{\bar a_1}\cdots \partial_{\bar a_r}F \right) \left( \partial_{b_1}\cdots \partial_{b_r}G \right), \end{split} \end{equation} where \(m(F\otimes G)=FG\) is ordinary pointwise multiplication. \end{definition} For \(G_J=I\), this is exactly the flat product of \cite[Example~1]{AlcaldeTyagi2023a}; the tensor \(\Sigma_J^{\bar ab}\) writes the same construction for an arbitrary positive Hermitian metric. The product was obtained there from the Kähler geometry of preferred observables. In the terminology later systematized by Karabegov~\cite{Karabegov1996}, it is a star product with separation of variables. The expansion \eqref{eq:star} is finite on polynomials. Associativity may be seen directly from the same holomorphic--antiholomorphic commutation that underlies the preferred-observable construction. Let \[ C := \Sigma_J^{\bar ab} \partial_{\bar a}\otimes\partial_b. \] On three tensor factors, denote by \(C_{ij}\) the operator \(C\) acting on the \(i\)-th and \(j\)-th factors. All derivatives have constant coefficients, and holomorphic and antiholomorphic derivatives commute. Hence \[ [C_{12},C_{13}] = [C_{12},C_{23}] = [C_{13},C_{23}] = 0. \] Both possible associations of the threefold product are therefore equal to \[ m^{(3)} \circ \exp\left[ \lambda \left( C_{12}+C_{13}+C_{23} \right) \right] (F\otimes G\otimes H). \] Thus \[ (F\st G)\st H = F\st(G\st H). \] The product is polarized by construction. If \(a\in\mathcal O_J\) is holomorphic, then \(\partial_{\bar a}a=0\); if \(b\in\overline{\mathcal O}_J\) is antiholomorphic, then \(\partial_ab=0\). Consequently, \begin{equation}\label{eq:separation-rule} a(z)\st F = a(z)F, \qquad F\st b(\wb) = Fb(\wb). \end{equation} Holomorphic functions act by ordinary multiplication from the left, and antiholomorphic functions act by ordinary multiplication from the right. Deformation occurs only in the complementary direction: an antiholomorphic dependence in the first factor contracts with a holomorphic dependence in the second. This rule is the flat global form of the preferred-observable construction. The coordinate functions make it explicit. Since they are linear, their star products terminate after the first derivative: \begin{equation}\label{eq:elementary} \begin{aligned} \wb^{\bar a}\st F &= \wb^{\bar a}F + \lambda \Sigma_J^{\bar ab}\partial_bF, \\ F\st z^b &= Fz^b + \lambda \Sigma_J^{\bar ab}\partial_{\bar a}F, \\ [\wb^{\bar a},z^b]_\star &:= \wb^{\bar a}\st z^b - z^b\st\wb^{\bar a} = \lambda\Sigma_J^{\bar ab}. \end{aligned} \end{equation} Thus the elementary commutator is the covariance contraction of the cell. The tensor \(\Sigma_J\) pairs an antiholomorphic covector in the first factor with a holomorphic covector in the second. More generally, \[ F\st G = FG + \lambda \Sigma_J^{\bar ab} \partial_{\bar a}F\, \partial_bG + O(\lambda^2). \] Antisymmetrization gives \begin{align*} [F,G]_\star &= \lambda \Sigma_J^{\bar ab} \left( \partial_{\bar a}F\,\partial_bG - \partial_{\bar a}G\,\partial_bF \right) + O(\lambda^2) \\ &= -i\lambda\{F,G\}_J + O(\lambda^2). \end{align*} Therefore \begin{equation}\label{eq:classical-limit} \lim_{\lambda\to0}F\st G = FG, \qquad \lim_{\lambda\to0} \frac{i}{\lambda}[F,G]_\star = \{F,G\}_J. \end{equation} The limit \(\lambda\to0\) removes the covariance scale and recovers the commutative product together with its Poisson bracket. The product also carries the natural real structure of the Kähler double. Define \[ F^*(\wb,z) := \overline{F(\bar z,\bar{\wb})}. \] This operation combines complex conjugation with interchange of the two factors. Since \(\Sigma_J\) is Hermitian, conjugation reverses the direction of every contraction, and hence \[ (F\st G)^* = G^*\st F^*. \] We have therefore obtained a unital \( * \)-algebra \[ \mathcal P_{J,\lambda} := (\Pol,\st,{}^*). \] The construction may be summarized as follows. The Kähler polarization separates the Hamiltonian vector field into holomorphic and antiholomorphic directions; the covariance tensor contracts those directions; and exponentiation of that contraction extends the preferred-observable rule to the full polynomial algebra: \[ (J,\Sigma_J) \longmapsto \Sigma_J^{\bar ab} \partial_{\bar a}\otimes\partial_b \longmapsto \star_{J,\lambda}. \] In this sense, the star product is determined directly by the positive polarization and its covariance cell.

\subsection{Symplectic and covariance contractions} The bidifferential contraction defining \(\st\) combines an antisymmetric Poisson part and a symmetric covariance part. This is the constant-coefficient ordering decomposition used by Yoshioka: star products with a common skew part represent the same Weyl algebra, while their symmetric parts specify different ordering expressions \cite{OMMY2007}. In the present setting, this algebraic decomposition has a direct statistical meaning. The fixed antisymmetric part is determined by \(\omega^{-1}\), while the symmetric part is determined by the covariance tensor of the cell. Set \[ C := \Sigma_J^{\bar ab} \partial_{\bar a}\otimes\partial_b. \] Under interchange of the two tensor factors, \(C\) decomposes as \[ C=C_{\mathrm s}+C_{\mathrm a}, \] where \begin{equation}\label{eq:CsCa} \begin{aligned} C_{\mathrm s} &:= \frac12\Sigma_J^{\bar ab} \left( \partial_{\bar a}\otimes\partial_b + \partial_b\otimes\partial_{\bar a} \right), \\ C_{\mathrm a} &:= \frac12\Sigma_J^{\bar ab} \left( \partial_{\bar a}\otimes\partial_b - \partial_b\otimes\partial_{\bar a} \right). \end{aligned} \end{equation} The two terms have different geometric roles. On the real form, \(C_{\mathrm a}\) is the contraction defined by the Poisson tensor \(\Pi=\omega^{-1}\), up to the normalization already fixed in \eqref{eq:poisson-double}. It depends only on the symplectic structure. By contrast, \(C_{\mathrm s}\) is the real covariance contraction associated with \(\Sigma_J\). It depends on the selected cell. Define the corresponding Moyal product by \begin{equation}\label{eq:moyal} F\stM G := m\circ e^{\lambda C_{\mathrm a}}(F\otimes G). \end{equation} Because \(C_{\mathrm a}\) depends only on \(\omega\), the Moyal product is independent of the covariance cell. It may be regarded as the neutral, symplectically invariant reference product. The passage from \(\stM\) to \(\st\) consists in adding the symmetric contraction \(C_{\mathrm s}\): \[ F\st G = m\circ e^{\lambda(C_{\mathrm a}+C_{\mathrm s})} (F\otimes G). \] Since the two contractions have constant coefficients, \[ [C_{\mathrm a},C_{\mathrm s}]=0. \] The separated product is therefore obtained from the Moyal product by a change of ordering generated by the covariance. Introduce the second-order operator \begin{equation}\label{eq:heat-operator} \Delta_J := \Sigma_J^{\bar ab} \partial_{\bar a}\partial_b, \qquad T_{J,\lambda} := \exp\left( \frac{\lambda}{2}\Delta_J \right). \end{equation} On the real form, \(\Delta_J\) is a positive constant-coefficient Laplacian determined by the covariance metric. For \(N=1\), \(G_J=1\), and \(z=(q+ip)/\sqrt2\), \[ \Delta_J = \partial_{\bar z}\partial_z = \frac12 \left( \partial_q^2+\partial_p^2 \right). \] Thus \(T_{J,\lambda}\) is a heat operator whose diffusion tensor is the covariance of the cell. 

\subsection{The product as Gaussian smoothing} The following theorem identifies the symmetric ordering transformation with Gaussian smoothing. It is the central link between the algebraic product and the statistical interpretation of the covariance cell. \begin{theorem}[The star product of a cell] \label{thm:smoothing} For all \(F,G\in\Pol\), \begin{equation}\label{eq:smoothing} T_{J,\lambda}\bigl(F\stM G\bigr) = T_{J,\lambda}F \;\st\; T_{J,\lambda}G. \end{equation} On the real form, \(T_{J,\lambda}\) is convolution with the centred Gaussian \(\rho_{J,\lambda/2}\) of covariance \[ \frac{\lambda}{2}\Sigma_J. \] Consequently, \[ T_{J,\lambda}^2 \] is convolution with the cell density \(\rho_{J,\lambda}\), whose covariance is \(\lambda\Sigma_J\). \end{theorem} \begin{proof} The Leibniz rule for \(\Delta_J\) gives \begin{equation}\label{eq:Delta-Leibniz} \Delta_J\circ m = m\circ \left( \Delta_J\otimes1 + 1\otimes\Delta_J + 2C_{\mathrm s} \right). \end{equation} Indeed, applying \(\Delta_J\) to a product gives \begin{align*} \Delta_J(FG) &= (\Delta_JF)G + F(\Delta_JG) \\ &\quad+ \Sigma_J^{\bar ab} \left( \partial_{\bar a}F\,\partial_bG + \partial_bF\,\partial_{\bar a}G \right), \end{align*} and the last line is precisely \(2m\circ C_{\mathrm s}(F\otimes G)\). All the operators involved have constant coefficients and therefore commute. Exponentiating \eqref{eq:Delta-Leibniz} gives, for every scalar \(s\), \begin{align} e^{s\Delta_J} \circ m\circ e^{\lambda C_{\mathrm a}} &= m\circ e^{s(\Delta_J\otimes1+1\otimes\Delta_J+2C_{\mathrm s})} e^{\lambda C_{\mathrm a}} \notag\\ &= m\circ e^{\lambda C_{\mathrm a}} e^{2sC_{\mathrm s}} \bigl( e^{s\Delta_J}\otimes e^{s\Delta_J} \bigr). \label{eq:heat-intertwining} \end{align} Choose \(s=\lambda/2\). Since \[ C=C_{\mathrm a}+C_{\mathrm s}, \] the right-hand side of \eqref{eq:heat-intertwining} becomes \[ m\circ e^{\lambda C} \bigl( T_{J,\lambda}\otimes T_{J,\lambda} \bigr). \] Applied to \(F\otimes G\), this is \[ T_{J,\lambda}F \st T_{J,\lambda}G. \] The left-hand side is \[ T_{J,\lambda}(F\stM G), \] which proves \eqref{eq:smoothing}. It remains to identify \(T_{J,\lambda}\) on the real form. For a positive symmetric tensor \(K\), the constant-coefficient heat operator \[ \exp\left( \frac12\langle K\partial,\partial\rangle \right) \] is convolution with the centred Gaussian of covariance \(K\). On polynomials, this identity may be verified directly by comparing the Taylor expansion with the Gaussian moment formula. After a complex-linear change of coordinates reducing \(G_J\) to the identity, the one-dimensional normalization above shows that \[ T_{J,\lambda} = \exp\left( \frac{\lambda}{2}\Delta_J \right) \] corresponds to covariance \[ \frac{\lambda}{2}\Sigma_J. \] Equivalently, \[ (T_{J,\lambda}F)(x) = \int_V F(x-y)\, \rho_{J,\lambda/2}(y)\,dy. \] Since the convolution of centred Gaussians adds their covariance tensors, \[ \rho_{J,\lambda/2}*\rho_{J,\lambda/2} = \rho_{J,\lambda}. \] It follows that \(T_{J,\lambda}^2\) is convolution with the Gaussian density of the full cell. \end{proof} \Cref{thm:smoothing} gives a precise meaning to the statement that \(\star_{J,\lambda}\) is the algebra of observables seen through the cell. In the terminology of~\cite{AlcaldeTyagi2023b}, \[ T_{J,\lambda}F \] is the statistical observable \[ F_\rho=\rho\circ F \] associated with the Gaussian prior \[ \rho=\rho_{J,\lambda/2}. \] Equation \eqref{eq:smoothing} states that the cell product is exactly the product for which passage to these statistical observables is multiplicative. This also gives a statistical interpretation of the ordering decomposition. The Moyal product is neutral: it retains only the antisymmetric contraction and therefore sees only the symplectic form. Choosing a covariance cell adds a positive symmetric contraction. Algebraically, this selects an ordering expression of the Moyal algebra. Statistically, it replaces a sharp observable by its Gaussian average at the resolution of half the cell: \[ F \longmapsto T_{J,\lambda}F = \rho_{J,\lambda/2}*F. \] The ordering and the prior are therefore two descriptions of the same datum. There is a useful reason for the factor \(1/2\). The product compares two observables, and each factor is smoothed by \(\rho_{J,\lambda/2}\). Their two smoothing covariances add: \[ \frac{\lambda}{2}\Sigma_J + \frac{\lambda}{2}\Sigma_J = \lambda\Sigma_J. \] Thus the Gaussian attached to each factor is the convolution square root of the full cell density. 

\subsection{Changing the cell} Let \(J_1\) and \(J_2\) define two balanced cells with the same symplectic scale \(\lambda\). Their star products have the same antisymmetric contraction, because both are associated with the same symplectic form \(\omega\). They differ only in their symmetric covariance contractions. There are two related, but distinct, ways in which such cells may be compared. A symplectic transformation may \emph{transport} one cell into the other, carrying its density and square-root amplitude with it. Alternatively, the two associated star products may be compared as different ordering expressions of the same Moyal algebra. The first is a dynamical comparison; the second is an ordering equivalence. The ordering-equivalence formula of Omori, Maeda, Miyazaki, and Yoshioka~\cite{OMMY2007} specializes in the present notation to \begin{equation}\label{eq:cell-intertwiner} \mathcal I_{J_2J_1} := T_{J_2,\lambda}T_{J_1,\lambda}^{-1} = \exp\left[ \frac{\lambda}{2} \left( \Delta_{J_2}-\Delta_{J_1} \right) \right]. \end{equation} This operator intertwines the two cell products: \begin{equation}\label{eq:cell-equivalence} \mathcal I_{J_2J_1} \bigl( F\star_{J_1,\lambda}G \bigr) = \mathcal I_{J_2J_1}F \star_{J_2,\lambda} \mathcal I_{J_2J_1}G. \end{equation} The equivalence operators satisfy \[ \mathcal I_{J_3J_2} \mathcal I_{J_2J_1} = \mathcal I_{J_3J_1}, \qquad \mathcal I_{J_2J_1}^{-1} = \mathcal I_{J_1J_2}. \] Thus the algebras associated with different cells are equivalent expressions of one Moyal algebra. Their symplectic commutation structure is the same; what changes is the statistical resolution through which observables are represented. \paragraph{Rotations and squeezes.} Suppose now that the second cell is obtained dynamically from the first. Let \[ S_t=e^{tA}\in\Sp(V,\omega), \qquad J_t=S_tJ_0S_t^{-1}, \qquad \Sigma_t=S_t\Sigma_0S_t^{\T}. \] The Cartan decomposition at \(J_0\), \[ A=A_{\mathfrak k}+A_{\mathfrak p}, \qquad A_{\mathfrak k}\in\mathfrak k_{J_0}, \quad A_{\mathfrak p}\in\mathfrak p_{J_0}, \] distinguishes two cases. If \(A\in\mathfrak k_{J_0}\), then \(S_t\) is \(J_0\)-unitary and \[ J_t=J_0, \qquad \Sigma_t=\Sigma_0. \] The Hamiltonian flow may rotate points within the cell, but it does not change the cell itself. Consequently, \[ \Delta_{J_t}=\Delta_{J_0}, \qquad T_{J_t,\lambda}=T_{J_0,\lambda}, \qquad \mathcal I_{J_tJ_0}=I. \] A \(J_0\)-unitary Hamiltonian therefore acts within one fixed statistical resolution and within one fixed cell algebra. If \(A\) has a nonzero \(\mathfrak p_{J_0}\)-component, then the flow squeezes the cell: \[ \dot\Sigma_t = A\Sigma_t+\Sigma_tA^{\T} \neq0 \] in general. The covariance, heat operator, and separated product all change: \[ \Sigma_0\longmapsto\Sigma_t, \qquad T_{J_0,\lambda}\longmapsto T_{J_t,\lambda}, \qquad \star_{J_0,\lambda} \longmapsto \star_{J_t,\lambda}. \] A squeeze is therefore not an internal symmetry of one cell algebra. It transports the entire resolution structure from one member of the family \[ \{\mathcal P_{J,\lambda}\}_{J\in\Jp} \] to another. This agrees with the square-amplitude picture. Let \[ \rho_{J,\lambda}(x) \propto \exp\left[ -\frac{1}{2\lambda} \langle G_Jx,x\rangle \right] \] denote the Gaussian density of the cell, with the normalization appropriate to the chosen real coordinates. Its positive square-root amplitude is \[ \psi_{J,\lambda} := \sqrt{\rho_{J,\lambda}}. \] Because \(S_t\) preserves Liouville volume, \begin{equation}\label{eq:cell-amplitude-transport} \rho_{J_t,\lambda} = \rho_{J_0,\lambda}\circ S_t^{-1}, \qquad \psi_{J_t,\lambda} = \psi_{J_0,\lambda}\circ S_t^{-1}. \end{equation} Thus Koopman transport gives a natural unitary comparison between the square-root amplitudes: \[ U_t\psi_{J_0,\lambda} = \psi_{J_t,\lambda}. \] For \(A\in\mathfrak k_{J_0}\), the amplitude is unchanged as a statistical cell: \[ \psi_{J_t,\lambda} = \psi_{J_0,\lambda}. \] For a squeeze, the amplitude changes shape but remains normalized. The Koopman operator does not compare two fixed Gaussian resolutions by deconvolution; it transports one Gaussian preparation into the other. It is therefore important to distinguish the two intertwiners: \[ \begin{aligned} U_t &:\ \text{transport of densities and amplitudes}, \\ \mathcal I_{J_tJ_0} &:\ \text{change of ordering expression at fixed coordinates}. \end{aligned} \] They describe the same change of covariance from complementary viewpoints, but they act on different objects. The first acts on classical amplitudes and is unitary. The second acts on observable symbols and is generally a mixture of smoothing and deconvolution. 

\begin{remark}[Comparing two cells requires a common refinement] \label{rem:refinement} Two distinct balanced cells of the same scale are never nested. Indeed, suppose that \[ \Sigma_{J_2}\geq\Sigma_{J_1} \] as positive tensors. Balanced cells at the same scale have equal determinant: 
\(\det\Sigma_J=1\) for every normalized balanced cell, so \(\det(\lambda\Sigma_J)=\lambda^{2N}\).

\[ \det\Sigma_{J_2} = \det\Sigma_{J_1}. \] Set \[ B := \Sigma_{J_1}^{-1/2} \Sigma_{J_2} \Sigma_{J_1}^{-1/2}. \] Then \[ B\geq I, \qquad \det B=1. \] Every eigenvalue of \(B\) is at least one, while their product is one. Hence every eigenvalue equals one, so that \[ B=I \qquad\text{and}\qquad \Sigma_{J_2}=\Sigma_{J_1}. \] Consequently, if \(J_1\neq J_2\), then \[ \Sigma_{J_2}-\Sigma_{J_1} \] is indefinite. The operator \[ \Delta_{J_2}-\Delta_{J_1} \] therefore diffuses in some directions and deconvolves in others. Accordingly, \[ T_{J_2,\lambda}T_{J_1,\lambda}^{-1} \] is never a pure smoothing operator unless the two cells coincide. This does not conflict with the unitary Koopman comparison \eqref{eq:cell-amplitude-transport}. The latter transports the coordinates, the covariance, and the amplitude together. By contrast, \(\mathcal I_{J_2J_1}\) compares two different resolutions after they have been expressed relative to the same underlying Moyal algebra. The Moyal product has no distinguished compatible complex structure: although a choice of \(J\) may be used to express its bidifferential operator in holomorphic and antiholomorphic coordinates, its antisymmetric contraction is the Poisson tensor \(\Pi=\omega^{-1}\) and is independent of that choice. It is therefore the common polarization-independent reference product for the family of cell products \(\{\star_{J,\lambda}\}_{J\in\Jp}\). In the language of statistical models, the Moyal algebra is the infinitely sharp reference through which the cell products are compared:

 \[ \begin{tikzcd}[column sep=large] & (\Pol,\star_M) \arrow[dl,"T_{J_1,\lambda}"'] \arrow[dr,"T_{J_2,\lambda}"] & \\ (\Pol,\star_{J_1,\lambda}) \arrow[rr,"\mathcal I_{J_2J_1}"'] && (\Pol,\star_{J_2,\lambda}). \end{tikzcd} \] Each cell algebra is obtained by smoothing the same neutral Moyal algebra. Passing directly between two distinct cells necessarily combines smoothing in some directions with deconvolution in others. \end{remark} \begin{remark}[Ordering as statistical resolution] In the general constant-coefficient theory, the symmetric part of the bidifferential contraction is an ordering parameter. For a positive polarization, the present construction identifies that parameter with a positive covariance tensor. Positivity makes the equivalence from the Moyal expression to one cell expression a heat operator and hence a genuine Gaussian smoothing: \[ T_{J,\lambda} = \exp\left( \frac{\lambda}{2}\Delta_J \right). \] The relative equivalence between two different cells need not itself be positive, because the difference of two positive covariance tensors need not be positive. The algebraic freedom of choosing an ordering is thereby identified with the geometric and statistical freedom of choosing a covariance cell: \[ \boxed{ \text{symmetric ordering} \ \longleftrightarrow\ \text{positive covariance} \ \longleftrightarrow\ \text{Gaussian resolution} }. \] The antisymmetric contraction encodes the symplectic structure. The symmetric contraction records the statistical resolution through which that structure is represented. \end{remark} The construction may be summarized by the two complementary chains \[ \begin{aligned} \omega &\longmapsto C_{\mathrm a} \longmapsto \star_M, \\ \Sigma_J &\longmapsto C_{\mathrm s} \longmapsto T_{J,\lambda} \longmapsto \star_{J,\lambda}, \end{aligned} \] and \[ \begin{aligned} S_t &:\ (\rho_{J_0,\lambda},\psi_{J_0,\lambda}) \longmapsto (\rho_{J_t,\lambda},\psi_{J_t,\lambda}), \\ \mathcal I_{J_tJ_0} &:\ \mathcal P_{J_0,\lambda} \longmapsto \mathcal P_{J_t,\lambda}. \end{aligned} \] The symplectic form fixes the commutation structure. The covariance cell fixes its separated expression and Gaussian resolution. \(J\)-unitary Hamiltonians preserve both; squeezing Hamiltonians transport them together to a new cell.
\section{Rotations act classically within a cell; squeezes do not} \label{sec:symmetry}

 A balanced covariance cell determines both a positive polarization and a star product. It is therefore natural to ask which classical Hamiltonian symmetries can be implemented as inner symmetries of the corresponding fixed-cell algebra \[ \mathcal P_{J,\lambda} = (\Pol,\star_{J,\lambda},{}^*). \] For an observable \(H\), the star commutator always defines an inner derivation, \[ \frac{i}{\lambda}\operatorname{ad}_\star(H) := \frac{i}{\lambda}[H,\cdot]_\star. \] The issue is whether this inner derivation agrees with the classical Hamiltonian derivation \[ \mathcal X_H:=\{H,\cdot\}_J. \] For affine Hamiltonians preserving the Kähler structure, the two derivations agree exactly. For general quadratic Hamiltonians, however, the answer depends on the Cartan decomposition \[ \sp(V,\omega)=\mathfrak k_J\oplus\mathfrak p_J. \] The compact part \(\mathfrak k_J\) preserves the covariance cell and is represented classically by inner derivations of its algebra. The noncompact part \(\mathfrak p_J\) changes the cell. Its classical action is therefore realized not inside one fixed algebra, but by isomorphisms between the algebras associated with different cells. This gives an algebraic formulation of the distinction between rotations and squeezes. Rotations are internal symmetries of a chosen statistical resolution; squeezes transport the resolution itself. The result develops the statistical-manifold and Cartan-decomposition picture introduced in~\cite{Alcalde2025}: the homogeneous space \(\Sp/\mathrm U(N)\) parametrizes the Gaussian resolution structures, while the present section determines precisely which Hamiltonian symmetries act internally on one fibre and which move between fibres. 

\subsection{Affine \(J\)-Hamiltonians} We begin with the infinitesimal symmetries that preserve the flat Kähler structure selected by \(J\). Let \[ Z_X = (Pz+u)^a\partial_a \] be an affine holomorphic vector field on \(V_J\), and let \[ X=Z_X+\overline{Z_X} \] be its real part. The linear part preserves the Hermitian metric \(G_J\) precisely when \[ P^\dagger G_J+G_JP=0, \] or equivalently \[ P\in\mathfrak u(V_J,g_J). \] Thus \(X\) generates an affine Kähler symmetry: its linear part rotates the cell into itself, while its translation part moves its centre without changing its shape. The Hamiltonian of \(X\), polarized to the Kähler double, is \begin{equation}\label{eq:hX} \begin{split} h_X(\wb,z) &= i\,\wb^{\T}G_JPz + i\,\wb^{\T}G_Ju - i\,z^{\T}G_J\bar u \\ &= i\bigl(Z_X\Phi_J-\chi_X\bigr), \qquad \chi_X(z):=z^{\T}G_J\bar u . \end{split} \end{equation} On the real form \(\wb=\bar z\), this function is real and satisfies \[ dh_X=\iota_X\omega. \] Equation \eqref{eq:hX} is the flat case of the Kähler-potential formula for preferred observables in \cite[Lemma~1]{AlcaldeTyagi2023a}. We call the functions \(h_X\) \emph{\(J\)-Hamiltonians}. Their homogeneous quadratic part, \[ h_P(\wb,z)=i\,\wb^{\T}G_JPz, \] has bidegree \((1,1)\). Under the quadratic comoment map, this sector is identified with \[ \mathfrak k_J\simeq\mathfrak u(V_J,g_J), \] the Lie algebra of linear transformations preserving the cell. \begin{proposition}[Exact inner action of affine \(J\)-Hamiltonians] \label{prop:affine} For every affine \(J\)-Hamiltonian \(h_X\) and every \(F\in\Pol\), \begin{equation}\label{eq:affine-inner} \frac{i}{\lambda}[h_X,F]_\star = \{h_X,F\}_J. \end{equation} Thus the classical Hamiltonian action of every affine Kähler symmetry is represented exactly by an inner derivation of \(\mathcal P_{J,\lambda}\). For two such Hamiltonians, \begin{equation}\label{eq:affine-extension} [h_X,h_Y]_\star = -i\lambda \bigl( h_{[X,Y]}+c(X,Y) \bigr), \end{equation} where \(c(X,Y)\) is a constant determined by the translation parts. For two pure translations, \[ c(X,Y)=\omega(u_X,u_Y). \] This is the usual Heisenberg central extension of the translation algebra. \end{proposition} \begin{proof} The Hamiltonian \(h_X\) is at most linear in the antiholomorphic variables \(\wb\) and at most linear in the holomorphic variables \(z\). Consequently, the expansion \eqref{eq:star} terminates after first order in both \[ h_X\star F \qquad\text{and}\qquad F\star h_X. \] Their difference is therefore \[ [h_X,F]_\star = -i\lambda\{h_X,F\}_J, \] which proves \eqref{eq:affine-inner}. Applying the same identity to \(F=h_Y\) shows that the nonconstant part of the commutator is the Hamiltonian of \([X,Y]\). Hamiltonians are defined only up to constants, and the translation terms produce the central cocycle \(c(X,Y)\). For pure translations, direct evaluation gives \[ c(X,Y)=\omega(u_X,u_Y). \] \end{proof} The proposition identifies the affine unitary group, together with its Heisenberg extension, as an exact classical inner symmetry of the fixed-cell algebra. The situation changes when the quadratic flow deforms the covariance cell. \subsection{General quadratic Hamiltonians} Let \(A\in\sp(V,\omega)\), and let \(H_A\) be the quadratic Hamiltonian defined by \[ \iota_{X_{H_A}}\omega=dH_A, \qquad X_{H_A}(x)=Ax. \] Relative to the complex structure \(J\), write the \((1,0)\)-part of the Hamiltonian vector field as \[ X_{H_A}^{1,0} = (P_Az+Q_A\wb)^a\partial_a. \] The coefficient \(P_A\) represents the component \(A_{\mathfrak k}\in\mathfrak k_J\), while \(Q_A\) represents \(A_{\mathfrak p}\in\mathfrak p_J\). Set \[ C_A:=iG_JP_A, \qquad B_A:=iG_JQ_A. \] The matrix \(C_A\) is Hermitian and \(B_A\) is complex symmetric. The polarized extension of \(H_A\) to the Kähler double then splits according to bidegree: \begin{equation}\label{eq:bidegree} H_A = \underbrace{ \wb^{\T}C_Az }_{ (1,1)\text{ sector: }\mathfrak k_J } + \underbrace{ \frac12z^{\T}B_Az + \frac12\wb^{\T}\bar B_A\wb }_{ (2,0)+(0,2)\text{ sectors: }\mathfrak p_J }. \end{equation} This is the polynomial counterpart of the Cartan decomposition. The mixed term generates a rotation preserving the cell. The pure holomorphic and antiholomorphic terms generate squeezes that change its shape and orientation. Under the quadratic comoment map \(A\mapsto H_A\), the Cartan commutation relations become \[ \{\mathcal H_{\mathfrak k}, \mathcal H_{\mathfrak k}\} \subseteq \mathcal H_{\mathfrak k}, \] \[ \{\mathcal H_{\mathfrak k}, \mathcal H_{\mathfrak p}\} \subseteq \mathcal H_{\mathfrak p}, \] and \[ \{\mathcal H_{\mathfrak p}, \mathcal H_{\mathfrak p}\} \subseteq \mathcal H_{\mathfrak k}. \] The \((1,1)\) Hamiltonians are linear in each set of polarized variables, so their star commutators terminate at first order. The pure \((2,0)\) and \((0,2)\) terms have nonzero second derivatives in \(z\) or \(\wb\). Their star commutators therefore contain ordering corrections. The smoothing equivalence of \cref{thm:smoothing} computes these corrections exactly. \begin{theorem}[Rotations are inner, squeezes are not] \label{thm:inner} Let \(A,B\in\sp(V,\omega)\), let \[ \mathcal X_A:=\{H_A,\cdot\}_J, \] and write \(T=T_{J,\lambda}\). Then \begin{equation}\label{eq:inner} \begin{split} \frac{i}{\lambda} \operatorname{ad}_\star(H_A) &= T\,\mathcal X_A\,T^{-1} \\ &= \mathcal X_A + \frac{\lambda}{2} [\Delta_J,\mathcal X_A], \end{split} \end{equation} and \begin{equation}\label{eq:cocycle} [H_A,H_B]_\star = -i\lambda \left( H_{[A,B]} + \frac{\lambda}{2} \Delta_JH_{[A,B]} \right). \end{equation} Consequently: \begin{enumerate}[label=(\alph*)] \item The classical Hamiltonian derivation is inner in the fixed-cell algebra, \[ \frac{i}{\lambda} \operatorname{ad}_\star(H_A) = \mathcal X_A \quad\text{on }\Pol, \] if and only if \[ A\in\mathfrak k_J. \] Equivalently, exact classical innerness holds if and only if the Hamiltonian flow preserves the covariance cell. \item The constants \[ \frac{\lambda}{2}\Delta_JH_{[A,B]} \] form a Lie-algebra coboundary. The shifted Hamiltonians \begin{equation}\label{eq:shifted-H} H_A^J := T_{J,\lambda}H_A = H_A+\frac{\lambda}{2}\Delta_JH_A \end{equation} satisfy \begin{equation}\label{eq:shifted-representation} [H_A^J,H_B^J]_\star = -i\lambda H_{[A,B]}^J \end{equation} exactly. 

\item The map \[ A \longmapsto \frac{i}{\lambda} \operatorname{ad}_\star(H_A) \] is a representation of \(\sp(V,\omega)\) by derivations of \(\mathcal P_{J,\lambda}\). For \(A\notin\mathfrak k_J\), however, the resulting inner derivation is the Gaussian conjugate \[ T_{J,\lambda}\mathcal X_AT_{J,\lambda}^{-1}, \] not the original classical derivation \(\mathcal X_A\). The scalar term in \eqref{eq:cocycle} disappears under \(\operatorname{ad}_\star\); hence the resulting derivations form an exact representation of \(\sp(V,\omega)\).

\end{enumerate} \end{theorem} 

\begin{proof} For a quadratic Hamiltonian \(H\), the Moyal commutator is exact: \begin{equation}\label{eq:moyal-quadratic-exact} \frac{i}{\lambda}[H,F]_{\star_M} = \{H,F\}_J. \end{equation} Indeed, only odd powers of the antisymmetric bidifferential contraction contribute to the commutator, and every odd term beyond the first contains at least three derivatives of \(H\). These terms vanish when \(H\) is quadratic. By the smoothing equivalence \eqref{eq:smoothing}, \[ T(F\star_M G) = TF\star TG. \] It follows that \[ \operatorname{ad}_\star(TH) = T\operatorname{ad}_{\star_M}(H)T^{-1}. \] For a quadratic Hamiltonian, \[ TH = H+\frac{\lambda}{2}\Delta_JH, \] and \(\Delta_JH\) is constant. It therefore commutes with every polynomial, so \[ \operatorname{ad}_\star(TH) = \operatorname{ad}_\star(H). \] Combining this observation with \eqref{eq:moyal-quadratic-exact} gives \[ \frac{i}{\lambda} \operatorname{ad}_\star(H_A) = T\mathcal X_AT^{-1}. \] Since \(\mathcal X_A\) is a linear vector field, \[ [\Delta_J,\mathcal X_A] \] is a constant-coefficient differential operator of order two. It commutes with \(\Delta_J\), and hence every higher commutator in the Baker--Campbell--Hausdorff expansion vanishes: \[ T\mathcal X_AT^{-1} = \mathcal X_A + \frac{\lambda}{2} [\Delta_J,\mathcal X_A]. \] This proves \eqref{eq:inner}. Applying \eqref{eq:inner} to \(H_B\), or equivalently using the exact Moyal relation and the smoothing intertwiner, gives \[ [H_A,H_B]_\star = -i\lambda \left( H_{[A,B]} + \frac{\lambda}{2}\Delta_JH_{[A,B]} \right), \] which is \eqref{eq:cocycle}. It remains to characterize when the correction vanishes. The commutator \[ [\Delta_J,\mathcal X_A] \] vanishes precisely when the linear flow generated by \(A\) preserves the covariance tensor \(\Sigma_J\): \[ A\Sigma_J+\Sigma_JA^{\T}=0. \] By \cref{prop:rotsqueeze}, this is equivalent to \[ A\in\mathfrak k_J. \] This proves part~(a). For part~(b), the shift \[ H_A\longmapsto H_A^J = H_A+\frac{\lambda}{2}\Delta_JH_A \] changes each Hamiltonian only by a constant. Substituting this shift into \eqref{eq:cocycle} absorbs the central term and yields \[ [H_A^J,H_B^J]_\star = -i\lambda H_{[A,B]}^J. \] Part~(c) follows from the Jacobi identity for the star commutator. \end{proof} The theorem contains two related but logically distinct statements. Part~(b) concerns the Lie-algebra representation carried by the quadratic Hamiltonians. Its scalar ordering correction is removable, as expected from Whitehead's lemma for the semisimple Lie algebra \(\sp(V,\omega)\). Part~(a) is stronger and is not removed by shifting Hamiltonians by constants. It characterizes when the inner derivation generated by the classical Hamiltonian is the classical Hamiltonian derivation itself: \[ \boxed{ \frac{i}{\lambda} \operatorname{ad}_\star(H_A) = \{H_A,\cdot\}_J \quad\Longleftrightarrow\quad A\in\mathfrak k_J }. \] Thus the maximal quadratic subalgebra acting classically by inner derivations of a fixed cell algebra is precisely the stabilizer algebra of that cell. In particular, the preferred-observable condition introduced in \cite{AlcaldeTyagi2023a}, \[ \frac{i}{\lambda}[H,F]_\star=\{H,F\}_J \qquad\text{for every }F, \] holds among quadratic Hamiltonians exactly for the \((1,1)\)-sector \(\mathcal H_{\mathfrak k}\). The preferred quadratic observables are therefore not selected by polynomial degree alone. They are selected geometrically as the Hamiltonians preserving the statistical resolution. \begin{example}[One symplectic plane] \label{ex:squeeze-commutator} Let \(N=1\), \(G_J=g>0\), and consider the two pure quadratic Hamiltonians \[ H^{0,2}:=\frac12\wb^2, \qquad H^{2,0}:=\frac12z^2. \] They belong to the squeeze sector \(\mathfrak p_J\). Direct calculation gives \[ [H^{0,2},H^{2,0}]_\star = \lambda g^{-1}\wb z + \frac{\lambda^2}{2}g^{-2}. \] The first term is the classical contribution: \[ \lambda g^{-1}\wb z = -i\lambda \{H^{0,2},H^{2,0}\}_J. \] The second is the scalar ordering correction predicted by \eqref{eq:cocycle}: \[ \frac{\lambda^2}{2}g^{-2} = -\frac{i\lambda^2}{2} \Delta_J \{H^{0,2},H^{2,0}\}_J. \] Although this constant can be absorbed into the shifted Hamiltonian, the second-order correction in the action \[ \frac{\lambda}{2} [\Delta_J,\mathcal X_A] \] cannot be removed by any scalar redefinition. The squeeze therefore does not act classically as an inner symmetry of the fixed-cell algebra. \end{example} The scalar term in \eqref{eq:cocycle} is the normal-ordering constant familiar from the metaplectic representation. The genuinely geometric obstruction is instead the failure of a squeeze to preserve the covariance tensor. A squeeze is a valid symplectic symmetry of the underlying phase space, but it is not an internal classical symmetry of an algebra whose product has been tied to one fixed cell. \subsection{Transport between cell algebras} Squeezes are not lost from the theory. They are implemented classically, but as transformations \emph{between} cell algebras rather than as classical inner transformations of a single one. Let \(S_t\in\Sp(V,\omega)\) be a symplectic flow and set \[ J_t:=S_tJ_0S_t^{-1}. \] Then \[ S_t:(V,J_0)\longrightarrow(V,J_t) \] is complex linear. It therefore induces a biholomorphic map of Kähler doubles \[ \widetilde S_t := S_t\times S_t: \mathbb D_{J_0} \longrightarrow \mathbb D_{J_t}. \] The precision and covariance tensors transform as \[ g_t=S_t^{-\T}g_0S_t^{-1}, \qquad \Sigma_t=S_t\Sigma_0S_t^{\T}, \] and the polarized Kähler potential is transported accordingly: \[ \Phi_{J_t}\circ\widetilde S_t = \Phi_{J_0}. \] Define the Koopman pullback \begin{equation}\label{eq:alpha} \alpha_{t,0}F := F\circ\widetilde S_t^{-1}. \end{equation} This map transports both observables and the polarization in which they are expressed. \begin{proposition}[Covariance of the cell algebras] \label{prop:alpha} The Koopman pullback \[ \alpha_{t,0}: \mathcal P_{J_0,\lambda} \longrightarrow \mathcal P_{J_t,\lambda} \] is an isomorphism of unital \( * \)-algebras. More precisely: \begin{enumerate}[label=(\roman*)] \item it intertwines the star products: \begin{equation}\label{eq:alpha-star} \alpha_{t,0} \bigl( F\star_{J_0,\lambda}G \bigr) = \alpha_{t,0}F \star_{J_t,\lambda} \alpha_{t,0}G; \end{equation} \item it intertwines Gaussian smoothing: \begin{equation}\label{eq:alpha-T} \alpha_{t,0}T_{J_0,\lambda} = T_{J_t,\lambda}\alpha_{t,0}; \end{equation} \item it maps \(J_0\)-holomorphic and \(J_0\)-antiholomorphic polynomials to \(J_t\)-holomorphic and \(J_t\)-antiholomorphic polynomials, respectively; \item it transports the elementary covariance relation to \begin{equation}\label{eq:transported-elementary} [\wb_t^{\bar a},z_t^b]_{\star_{J_t,\lambda}} = \lambda\Sigma_t^{\bar ab}. \end{equation} \end{enumerate} \end{proposition} \begin{proof} By the chain rule, derivatives of a pulled-back function transform by \(S_t^{-1}\). At the same time, the covariance tensor transforms by the congruence \[ \Sigma_t=S_t\Sigma_0S_t^{\T}. \] In every contracted pair of derivatives in \eqref{eq:star}, the factors of \(S_t^{-1}\) are cancelled by the factors of \(S_t\) carried by \(\Sigma_t\). Each term in the star-product expansion is therefore covariant, which proves \eqref{eq:alpha-star}. The same cancellation gives \[ \alpha_{t,0}\Delta_{J_0} = \Delta_{J_t}\alpha_{t,0}. \] Exponentiating this identity yields \eqref{eq:alpha-T}. Complex linearity of \(S_t:(V,J_0)\to(V,J_t)\) proves the preservation of holomorphic type, and applying \eqref{eq:alpha-star} to the coordinate functions gives \eqref{eq:transported-elementary}. Compatibility with the involution and preservation of the unit follow directly from the definition of the pullback. \end{proof} The covariance statement may be summarized by the commutative diagram \[ \begin{tikzcd}[column sep=large] \mathcal P_{J_0,\lambda} \times \mathcal P_{J_0,\lambda} \arrow[r,"\star_{J_0,\lambda}"] \arrow[d,"\alpha_{t,0}\times\alpha_{t,0}"'] & \mathcal P_{J_0,\lambda} \arrow[d,"\alpha_{t,0}"] \\ \mathcal P_{J_t,\lambda} \times \mathcal P_{J_t,\lambda} \arrow[r,"\star_{J_t,\lambda}"] & \mathcal P_{J_t,\lambda}. \end{tikzcd} \] The two cases in \cref{thm:inner} now have a unified interpretation. If \[ A\in\mathfrak k_{J_0}, \] then \(J_t=J_0\) and the covariance cell does not move. The maps \(\alpha_{t,0}\) form a one-parameter group of automorphisms of the single algebra \(\mathcal P_{J_0,\lambda}\). By \cref{thm:inner}(a), this group is inner: \[ \alpha_t(F) = E_A(t)\star F\star E_A(-t), \] where, formally, \[ E_A(t) := \operatorname{Exp}_\star \left( \frac{it}{\lambda}H_A \right). \] Rotations are therefore internal symmetries of one cell algebra. If \(A\) has a nonzero \(\mathfrak p_{J_0}\)-component, then \[ J_t\neq J_0 \] in general. The pullback \(\alpha_{t,0}\) relates distinct members of the family \[ \left\{ \mathcal P_{J,\lambda} \right\}_{J\in\Jp}. \] It remains a classical symplectic transport, but it is not an automorphism of any one fixed-cell algebra. By \cref{thm:inner}(a), no addition of scalar ordering constants can turn its infinitesimal action into the classical inner derivation of that fixed algebra. The natural algebraic object is therefore not a single star algebra, but the family \begin{equation}\label{eq:algebra-bundle} \bigsqcup_{J\in\Jp} \mathcal P_{J,\lambda} \longrightarrow \Jp. \end{equation} The compact directions \(\mathfrak k_J\) act vertically, by inner symmetries within a fibre. The noncompact directions \(\mathfrak p_J\simeq T_J\Jp\) act horizontally, by transporting one cell algebra to another: \[ \boxed{ \begin{aligned} \mathfrak k_J &: &&\text{inner symmetry within }\mathcal P_{J,\lambda}, \\ \mathfrak p_J &: &&\text{covariant transport between distinct cell algebras}. \end{aligned} } \] This is the algebraic counterpart of the Cartan geometry of covariance cells. Rotations preserve the selected statistical resolution; squeezes move that resolution through its statistical manifold.

\section{Star polarization and the Fock module of a cell}\label{sec:polarization}

The algebra $\mathcal P_{J,\lambda}$ does not yet say on what it acts.
Following Fronsdal~\cite{Fronsdal1979} (see also~\cite[\S6]{AlcaldeTyagi2023a}), we impose the polarization of the cell as equations \emph{inside} the algebra and show that their solutions form a module whose cyclic element is the Gaussian density of the cell.

\subsection{Right polarization, left transport}

The functions $\ell_{\bar c}:=\partial_{\bar c}\Phi_J=(G_Jz)_{\bar c}$ are linear, Poisson commute, and their Hamiltonian vector fields span the antiholomorphic directions $\partial_{\bar a}$, i.e.\ the polarization $\overline{L_J}$.
Since $\ell_{\bar c}\in\mathcal O_J$, they also star commute: $\ell_{\bar c}\st\ell_{\bar d}=\ell_{\bar c}\ell_{\bar d}$.

Gaussians are not polynomials, so we enlarge the space on which the equations are posed to
\[
\mathcal E_{J,\lambda}:=\bigl\{e^{-\Phi_J/\lambda+\langle\wb,\eta\rangle}\,p\ :\ p\in\Pol,\ \eta\in\C^N\bigr\}.
\]
For $F\in\Pol$ and $\Psi\in\mathcal E_{J,\lambda}$, both $F\st\Psi$ and $\Psi\st\ell_{\bar c}$ are given by finite sums in \eqref{eq:star} (the series is limited by the $\wb$-degree of $F$, resp.\ the $z$-degree of $\ell$), lie in $\mathcal E_{J,\lambda}$, and are associative in every triple $(F,\Psi,\ell)$.

\begin{definition}
For $\chi\in\C^N$, the \emph{star-polarized sector} of the cell is
\[
\mathfrak K^\chi_{J,\lambda}:=\{\Psi\in\mathcal E_{J,\lambda}:\ \Psi\st\ell_{\bar c}=\chi_{\bar c}\Psi,\ c=1,\dots,N\}.
\]
\end{definition}

The equations are compatible because the $\ell_{\bar c}$ star commute.
Placing the polarization on the right and transport on the left gives stability for free:

\begin{proposition}[Left stability]\label{prop:leftstable}
If $\Psi\in\mathfrak K^\chi_{J,\lambda}$ and $F\in\mathcal P_{J,\lambda}$, then $F\st\Psi\in\mathfrak K^\chi_{J,\lambda}$. Hence $\mathfrak K^\chi_{J,\lambda}$ is a left $\mathcal P_{J,\lambda}$-module.
\end{proposition}

\begin{proof}
$(F\st\Psi)\st\ell_{\bar c}=F\st(\Psi\st\ell_{\bar c})=\chi_{\bar c}\,F\st\Psi$.
\end{proof}

No normalizer condition on $F$ is needed: left and right multiplications commute by associativity.

\subsection{The Gaussian cell as cyclic element}

By \eqref{eq:elementary}, $\Psi\st\ell_{\bar c}=(G_Jz)_{\bar c}\Psi+\lambda\,\partial_{\bar c}\Psi$, so the polarization equations are the first-order system
\begin{equation}\label{eq:polar-eq}
\lambda\,\partial_{\bar c}\Psi+\bigl((G_Jz)_{\bar c}-\chi_{\bar c}\bigr)\Psi=0
\iff
\partial_{\bar c}\Bigl(e^{(\Phi_J-\langle\wb,\chi\rangle)/\lambda}\,\Psi\Bigr)=0.
\end{equation}

\begin{theorem}[The Gaussian cell]\label{thm:gaussian}
$\mathfrak K^\chi_{J,\lambda}=\bigl\{a(z)\,e^{-(\Phi_J-\langle\wb,\chi\rangle)/\lambda}:\ a\in\C[z]\bigr\}$.
For $\chi=0$ the cyclic element $\Omega_{J,\lambda}:=e^{-\Phi_J/\lambda}$ restricts on the real form to $e^{-\phi_J/\lambda}$, which, normalized, is the Gaussian density $\rho_{J,\lambda}$ of covariance $\lambda\Sigma_J$:
\[
\rho_{J,\lambda}=\frac{\det G_J}{(\pi\lambda)^N}\,e^{-\bar z^{\T}G_Jz/\lambda},\qquad
\int z^a\rho_{J,\lambda}=0,\qquad\int z^a\bar z^{\bar b}\rho_{J,\lambda}=\lambda\Sigma_J^{\bar ba}.
\]
For $\chi\ne0$, up to a holomorphic factor, the restriction is the same Gaussian centred at $z_0=\Sigma_J\chi$: a translated cell.
\end{theorem}

\begin{proof}
By \eqref{eq:polar-eq}, $e^{(\Phi_J-\langle\wb,\chi\rangle)/\lambda}\Psi$ is annihilated by every $\partial_{\bar c}$ and lies in $e^{\langle\wb,\eta\rangle}\Pol$, hence is a polynomial in $z$.
The moments are the standard complex Gaussian integrals (\cref{app:gauss}).
For the displaced case, complete the square: $\bar z^{\T}G_Jz-\bar z^{\T}G_Jz_0-\bar z_0^{\T}G_Jz=(z-z_0)^\dagger G_J(z-z_0)-\text{const}$, and $e^{\bar z_0^{\T}G_Jz/\lambda}$ is holomorphic.
\end{proof}

\begin{remark}[Density, amplitude, and the factor $\frac12$]\label{rem:half}
The cyclic element selected by the polarization at scale $\lambda$ is the \emph{density} of the cell, not its amplitude.
The amplitude $\sqrt{\rho_{J,\lambda}}\propto e^{-\phi_J/2\lambda}$ satisfies the equation with coefficient $\frac12$, $\bigl(\lambda\partial_{\bar c}+\frac12\partial_{\bar c}\phi_J\bigr)\psi=0$, which is the polarization of the algebra at scale $2\lambda$.
\Cref{thm:fock} below shows how the two readings fit together: module elements $a\,\Omega_{J,\lambda}$ correspond to amplitudes $a\sqrt{\rho_{J,\lambda}}$.
\end{remark}

\subsection{The Fock module}

\begin{theorem}[Fock module of a cell]\label{thm:fock}
On $\mathfrak K^0_{J,\lambda}=\mathcal O_J\,\Omega_{J,\lambda}$ the left action of $\mathcal P_{J,\lambda}$ is
\begin{equation}\label{eq:fock}
z^b\st(a\,\Omega_{J,\lambda})=(z^ba)\,\Omega_{J,\lambda},\qquad
\wb^{\bar a}\st(a\,\Omega_{J,\lambda})=\bigl(\lambda\Sigma_J^{\bar ab}\partial_ba\bigr)\,\Omega_{J,\lambda},
\end{equation}
the Fock representation of $[\wb^{\bar a},z^b]=\lambda\Sigma_J^{\bar ab}$ on $\C[z]$ with vacuum $\Omega_{J,\lambda}$.
With the inner product
\[
\langle a\,\Omega_{J,\lambda},\,b\,\Omega_{J,\lambda}\rangle:=\int_{\C^N}\overline{a(z)}\,b(z)\,\rho_{J,\lambda}(z)\,d^{2N}z,
\]
it is a $*$-representation: the action of $F^*$ is the adjoint of the action of $F$.
Its completion is the Fock space $\Fock$ of holomorphic functions square-integrable against the cell density~\cite{Bargmann1961}, and
\[
\Fock\longrightarrow L^2(V,\vol),\qquad a\,\Omega_{J,\lambda}\longmapsto a\sqrt{\rho_{J,\lambda}},
\]
is an isometry onto a closed subspace of classical Koopman amplitudes.
\end{theorem}

\begin{proof}
The first formula is the separation rule.
For the second, $\wb^{\bar a}\st(a\Omega)=\wb^{\bar a}a\Omega+\lambda\Sigma_J^{\bar ab}\partial_b(a\Omega)$ and $\lambda\Sigma_J^{\bar ab}\partial_b\Omega=-\wb^{\bar a}\Omega$.
For the adjoint, integrate by parts against $\rho_{J,\lambda}$: since $\partial_b\rho_{J,\lambda}=-\lambda^{-1}(\bar z^{\T}G_J)_b\rho_{J,\lambda}$ and $\bar a$ is antiholomorphic, multiplication by $z^b$ and $\lambda\Sigma_J^{\bar ab}\partial_b$ are mutually adjoint.
The isometry is $\int|a|^2\rho_{J,\lambda}=\|a\sqrt{\rho_{J,\lambda}}\|^2$.
\end{proof}

The holomorphic factor carries the variable part of the amplitude and the Gaussian carries the fixed cell:
\[
\text{amplitude}=\underbrace{a(z)}_{\text{polarized amplitude}}\ \times\ \underbrace{\sqrt{\rho_{J,\lambda}}}_{\text{covariance cell}} .
\]
The ``polarized Fock realization'' is thus not an additional quantum structure: it is the subspace of classical amplitudes that are holomorphic multiples of the amplitude of one cell, on which the cell algebra acts by creation (multiplication by $z$) and annihilation (derivation by $\lambda\Sigma_J\partial$).

\subsection{Koopman transport of Fock spaces}

\begin{theorem}[Koopman transport moves Fock spaces]\label{thm:koopmanfock}
Let $U_t\psi=\psi\circ S_t^{-1}$ be the Koopman operator of $H_A$ on $L^2(V,\vol)$. Then
\[
U_t\bigl(a\sqrt{\rho_{J_0,\lambda}}\bigr)=(a\circ S_t^{-1})\sqrt{\rho_{J_t,\lambda}},
\]
with $a\circ S_t^{-1}$ $J_t$-holomorphic. Hence $U_t$ maps $\mathcal F_{J_0,\lambda}$ unitarily onto $\mathcal F_{J_t,\lambda}$, intertwining the module actions through $\alpha_{t,0}$.
For $A\in\mathfrak k_{J_0}$ it preserves $\mathcal F_{J_0,\lambda}$ and restricts there to the unitary action of $\mathrm U(V,J_0)$; for $A\notin\mathfrak k_{J_0}$ it moves the Fock space of one cell onto that of another.
\end{theorem}

\begin{proof}
$\rho_{J_0,\lambda}\circ S_t^{-1}=\rho_{J_t,\lambda}$ by \eqref{eq:transport}, and $d(a\circ S_t^{-1})\circ J_t=da\circ J_0\circ S_t^{-1}=i\,d(a\circ S_t^{-1})$.
The intertwining follows from \cref{prop:alpha} and the left-stability argument.
\end{proof}

\Cref{thm:koopmanfock} replaces the metaplectic action used in~\cite{Alcalde2025} by its classical counterpart: a squeeze does not act within the Fock space of a cell but maps it onto the Fock space of the squeezed cell, and on amplitudes it is nothing but the Koopman operator of a linear symplectic flow.
Together, \cref{thm:smoothing,thm:inner,thm:gaussian,thm:fock,thm:koopmanfock} say that the positive polarization, the cell algebra and the classical Gaussian amplitude are not three structures but one, transported as a whole by the symplectic group.

\section{Conclusion and outlook}\label{sec:outlook} A localized classical preparation carries more structure than its phase-space centre. For the balanced Gaussian preparations studied here, its covariance tensor, precision metric, compatible complex structure, and positive Lagrangian polarization are equivalent descriptions of one object: \[ \Sigma_J \longleftrightarrow g_J \longleftrightarrow J \longleftrightarrow L_J. \] Their parameter space at fixed resolution scale is the Siegel space \[ \Jp\simeq\Sp(V,\omega)/\mathrm U(V,\omega,J_0). \] This identification organizes the results of the paper. A symplectic connection transfers the linearized dynamics along a motion to one fixed tangent fibre, where covariance evolves according to \[ \Sigma_t=S_t\Sigma_0S_t^{\T}. \] The same cell determines a separated star product. Its antisymmetric part is fixed by the symplectic form, while its symmetric part is fixed by the covariance. The equivalence \[ T_{J,\lambda}(F\star_M G) = T_{J,\lambda}F \star_{J,\lambda} T_{J,\lambda}G \] shows that the change from Moyal to cell ordering is Gaussian smoothing. The Cartan splitting \[ \sp(V,\omega) = \mathfrak k_J\oplus\mathfrak p_J \] then separates two kinds of dynamics. The \(\mathfrak k_J\)-directions preserve the cell and act classically within its algebra. The \(\mathfrak p_J\)-directions squeeze the cell and transport observables between different cell algebras. Fronsdal's star-polarization equations extract from each algebra a Fock module whose cyclic element is the Gaussian density of the cell. Its square-root realization is a closed subspace of classical Koopman amplitudes, transported unitarily with the cell. Thus the main correspondence is \[ \boxed{ \begin{aligned} \text{balanced covariance cell} &\longleftrightarrow \text{positive polarization} \longleftrightarrow \text{Gaussian smoothing} \\ &\longleftrightarrow \text{cell star algebra} \longleftrightarrow \text{star-polarized Fock module}. \end{aligned}} \] No Planck constant is required: \(\lambda\) is a classical resolution scale, and measured observables remain commuting multiplication operators. \paragraph{Two transports.} The family of cell algebras over \(\Jp\) carries two distinct transports. The dynamical map \[ \alpha_{t,0}: \mathcal P_{J_0,\lambda} \longrightarrow \mathcal P_{J_t,\lambda} \] moves observables together with the cell and corresponds, on square-root amplitudes, to unitary Koopman transport. By contrast, \[ \mathcal I_{J_2J_1} = T_{J_2,\lambda}T_{J_1,\lambda}^{-1} \] compares two ordering expressions through their common Moyal algebra. For distinct cells at the same scale, this relative equivalence smooths in some directions and deconvolves in others. The square-amplitude picture also supplies a natural comparison between translated cells. For two cells with the same \(J\), scale \(\lambda\), and centres \(z_0,z_1\), \[ \left\langle \sqrt{\rho_{z_0}}, \sqrt{\rho_{z_1}} \right\rangle = \exp\left[ -\frac{1}{4\lambda} (z_0-z_1)^\dagger G_J(z_0-z_1) \right]. \] Hence \[ -4\lambda \log \left\langle \sqrt{\rho_{z_0}}, \sqrt{\rho_{z_1}} \right\rangle = D_J(z_0,z_1). \] The quadratic Kähler separation is therefore recovered directly from the statistical overlap of the cell amplitudes. \paragraph{Outlook.} Three extensions appear natural. \begin{enumerate}[label=(\roman*)] \item \emph{General covariances.} An unbalanced covariance carries one Williamson scale for each symplectic plane. Extending the cell algebras and polarized modules to this case requires separating polarization data from the additional relative scales. \item \emph{Information geometry and curvature.} For cells with different centres and covariances, the amplitude overlap combines displacement with shape mismatch. Its relation to the invariant metric and dual connections on the space of cells remains to be developed. On curved Kähler manifolds, the constant-coefficient heat operator must be replaced by a geometric heat kernel with curvature corrections. \item \emph{Resolution and thermal relaxation.} The resolution scale \(\lambda\) should be distinguished from the thermal scale \(\beta^{-1}=k_BT\). Hamiltonian dynamics transports cells, whereas metriplectic friction and diffusion may drive them toward a thermal covariance. Their joint treatment should describe how a finite-resolution preparation is transported and ultimately relaxes toward equilibrium. \end{enumerate} The flat balanced theory developed here is the case in which covariance, polarization, ordering, and amplitude can be identified exactly. Its extension to general covariances, curved geometry, and thermal relaxation is the natural next step.

\section*{AI Collaboration Statement} Generative AI tools were used during the preparation of this manuscript as editorial and expository aids. Their assistance included rewriting and reorganizing passages, improving the clarity of mathematical arguments, reformulating or expanding proofs and explanatory passages from material and directions supplied by the author, suggesting terminology, and checking notational and presentational consistency. The research program, mathematical constructions, results, interpretations, historical claims, and conclusions are those of the author. AI-generated suggestions were treated as provisional text and were critically reviewed, revised, and, where necessary, corrected by the author. The author takes full responsibility for the accuracy, originality, citations, and scientific content of the manuscript.
\appendix
\section{Sign conventions}\label{app:signs}

\begin{center}\small
\begin{tabular}{@{}ll@{}}
\toprule
Hamiltonian vector field & $\iota_{X_H}\omega=dH$ \\
Poisson bracket & $\{F,G\}=\omega(X_F,X_G)$, $X_G(F)=\{F,G\}$ \\
Quadratic Hamiltonian & $H_A(x)=-\frac12\omega(x,Ax)$, $X_{H_A}=Ax$, $\{H_A,H_B\}=H_{[A,B]}$ \\
Kähler data & $g_J(u,v)=\omega(u,Jv)$, $\Sigma_J=g_J^{-1}$, $\omega=i\,g_{a\bar b}\,dz^a\wedge d\bar z^{\bar b}$ \\
Bracket on the double & $\{F,G\}_J=-i\Sigma_J^{\bar ba}(\partial_aF\partial_{\bar b}G-\partial_{\bar b}F\partial_aG)$ \\
Star commutator & $[F,G]_\star=-i\lambda\{F,G\}_J+O(\lambda^2)$ \\
Koopman & $U_t=e^{-it\Koop}$, $\Koop=-iX_H$ \\
\bottomrule
\end{tabular}
\end{center}
Reversing the convention $\iota_{X_H}\omega=dH$ reverses the sign of every bracket and of $H_A$.

\section{Gaussian normalizations}\label{app:gauss}

\paragraph{Real form.} On $V\simeq\R^{2N}$ the Gaussian of covariance $\Sigma$ is $(2\pi)^{-N}(\det\Sigma)^{-1/2}e^{-\frac12x^{\T}\Sigma^{-1}x}$, with amplitude $(2\pi)^{-N/2}(\det\Sigma)^{-1/4}e^{-\frac14x^{\T}\Sigma^{-1}x}$ and characteristic function $e^{-\frac12\xi^{\T}\Sigma\xi}$.
On the doubled tangent space $W$ ($\dim_\R W=4N$) the constants are $(2\pi)^{-2N}(\det\Sigma)^{-1/2}$ and $(2\pi)^{-N}(\det\Sigma)^{-1/4}$.

\paragraph{Complex form.} $\int_{\C^N}e^{-\bar z^{\T}G_Jz/\lambda}\,d^{2N}z=(\pi\lambda)^N/\det G_J$, whence the normalization of $\rho_{J,\lambda}$ in \cref{thm:gaussian}; its moments are $\int z^a\bar z^{\bar b}\rho_{J,\lambda}=\lambda\Sigma_J^{\bar ba}$.
The normalized amplitude is $\sqrt{\rho_{J,\lambda}}=(\det G_J)^{1/2}(\pi\lambda)^{-N/2}e^{-\bar z^{\T}G_Jz/2\lambda}$.

\paragraph{Symplectic Fourier transform.} With $(\mathcal F_\omega f)(u)=(2\pi)^{-N}\int_Ve^{-i\omega(u,y)}f(y)\,dy$, the Gaussian $e^{-\frac12\langle\Sigma^{-1}y,y\rangle}$ is mapped to $(\det\Sigma)^{1/2}e^{-\frac12\langle\omega^\flat u,\Sigma\,\omega^\flat u\rangle}$, realizing the precision--covariance exchange of \cref{prop:threefaces} intrinsically.


\end{document}